\documentclass[10pt, journal]{IEEEtran}  % Comment this line out
\IEEEoverridecommandlockouts                              % This command is only
\usepackage[utf8]{inputenc}
\usepackage[T1]{fontenc}

\usepackage{graphics} % for pdf, bitmapped graphics files
\usepackage{epsfig} % for postscript graphics files
\usepackage{amsmath,amssymb,amsthm}
\usepackage{algorithm2e}
\usepackage{cite}
\usepackage{subfigure}
\usepackage{multirow}
\usepackage{makecell}
\usepackage{threeparttable}
\usepackage{bbding}
\usepackage{enumerate}

\title{\LARGE \bf
Environment-independent mmWave Fall Detection with Interacting Multiple Model
} 

\author{ \parbox{3 in}{\centering Xuyao Yu\\
        IoT Thrust, Information Hub\\
        The Hong Kong University of Science and Technology (Guangzhou)\\
        Guangzhou, China\\
        {\tt\small xyu991@connect.hkust-gz.edu.cn}}
        \hspace*{ 0.5 in}
        \parbox{3 in}{ \centering Jiazhao Wang \\
        School of Information Systems Technology and Design \\
        Singapore University of Technology and Design\\
        Singapore\\
        {\tt\small jiazhao\_wang@mymail.sutd.edu.sg}}\\
        \parbox{3 in}{ \centering Wenchao Jiang \\
        School of Information Systems Technology and Design \\
        Singapore University of Technology and Design\\
        Singapore\\
        {\tt\small wenchao\_jiang@sutd.edu.sg}}
}

\begin{document}

\maketitle
\thispagestyle{empty}
\pagestyle{empty}

%%%%%%%%%%%%%%%%%%%%%%%%%%%%%%%%%%%%%%%%%%%%%%%%%%%%%%%%%%%%%%%%%%%%%%%%%%%%%%%%
\begin{abstract}

    The ageing society brings attention to daily elderly care through sensing technologies. The future smart home is expected to enable in-home daily monitoring, such as fall detection, for seniors in a non-invasive, non-cooperative, and non-contact manner. The mmWave radar is a promising candidate technology for its privacy-preserving and non-contact manner. However, existing solutions suffer from low accuracy and robustness due to environment dependent features. In this paper, we present FADE (\underline{FA}ll \underline{DE}tection), a practical fall detection radar system with enhanced accuracy and robustness in real-world scenarios. The key enabler underlying FADE is an interacting multiple model (IMM) state estimator that can extract environment-independent features for highly accurate and instantaneous fall detection. Furthermore, we proposed a robust multiple-user tracking system to deal with noises from the environment and other human bodies. We deployed our algorithm on low computing power and low power consumption system-on-chip (SoC) composed of data front end, DSP, and ARM processor, and tested its performance in real-world. The experiment shows that the accuracy of fall detection is up to 95\%. 

\end{abstract}

%%%%%%%%%%%%%%%%%%%%%%%%%%%%%%%%%%%%%%%%%%%%%%%%%%%%%%%%%%%%%%%%%%%%%%%%%%%%%%%%
\section{Introduction}

The global ageing population calls for smart health care. In 2020, the global statistic for people over 65 years old is 702 million, accounting for 9.1\% of the global population, and this number is estimated to reach 1.5 billion by 2050, accounting for 15.9\% of the global population\cite{UnitedNations2019}. Of this elderly population, approximately 28-35\% fall each year\cite{CAMPBELL1981}.  What makes it worse is that falling is also a signal for some severe diseases. Therefore, immediate fall detection can save lives and provide instant information feedback to family members or institutions. Research on fall detection has also sprung up, looking for innovative methods to bring prompt rescue to the elderly.

Falls are commonly defined as \textit{inadvertently coming to rest on the ground, floor, or other lower level, excluding intentional change in position to rest in furniture, wall or other objects}\cite{WHO2007}. The mainstream solutions to extract people's motion and posture features in the process of falling include the wearable-based\cite{Seketa2021,Lee2015}, vision-based\cite{Stone2015,Nunez-Marcos2017,Rougier2011,Auvinet2011}, and radar-based approaches\cite{Tomii2012,Karsmakers2012,Wu2013,Su2015,Sadreazami2019,Jin2019,Takabatake2019,Wang2020,Ma2020,Hanifi2021,Maitre2021,Jin2019,Jin2020}. The wearable-based solutions are most widely used and most technologically mature at this stage. They commonly use inertial sensors which can provide accurate 3-axis acceleration and angular velocity information. The shortcomings of these approaches are that they are intrusive, easily broken, and must be worn or carried. The vision-based solutions become popular with the success of deep learning technology in the computer vision field. However, vision-based fall detection algorithms are born with high requirements of computing power. Meanwhile, the privacy issues caused by the camera are unavoidable, making the user acceptance of vision-based solutions very low \cite{Wang2020a}. 

In contrast to the two aforementioned approaches, the radar-based approaches, such as the mmWave radars, can achieve non-invasive and non-contact fall detection in a device-free manner. It is especially suitable for the care of elderly living alone or in the nursing house. Technically, the radar-based approaches are based on the intervention of the human body on the propagation of the radar signals. By processing the received signal, characteristics of the target can be obtained in the form of the point cloud. However, due to the complex and diverse deployment scenarios, the accuracy and robustness are the main challenges for the radar-based approaches to be widely applied.

In this paper, we propose FADE, a practical radar-based fall detection system with high accuracy and robustness. It is achieved by identifying stages of falling and distinguish a falling event from Activities of Daily Living (ADL) based on the most characteristic stage. Specifically, we adopt the interacting multiple models (IMM) algorithm \cite{Blom1988} to identify the stages of falling and their corresponding state features which are environment-independent and used to be captured by wearable accelerometers \cite{Seketa2021}. Then we focus only on the stage where the centroid of the human body declines sharply for fall detection. In such a manner, FADE can successfully identify the subtle differences between the falling activity and some similar ADL, such as the sit-down activity. In addition, FADE achieves higher robustness by well filtering out noise signals from indoor scatterers, such as interior walls, furniture, and another human body. 

The technical contributions of the paper are listed as follows:
\begin{itemize}
  \item FADE contributes an accurate fall detector based on stages of falling through the IMM algorithm to extract environment-independent features and to separate ADL and fall, which increased system robustness and greatly lowers the false alarm rate.
  \item FADE induces a tracking and denoising system to enable ghost target cancellation and multi-user fall detection.
  \item We implement FADE on the commodity off-the-shelf mmWave radar system-on-chip (SoC) and thoroughly evaluate its performance in accuracy, robustness, and time consumption. Our experimental results show that FADE outperforms SoAs and reaches the accuracy of fall detection up to 95\%.
\end{itemize}

\section{Related Work}

In the research field of radar-based fall detection, most works are based on micro-Doppler features\cite{Tomii2012,Karsmakers2012,Wu2013,Su2015,Sadreazami2019,Jin2019,Takabatake2019,Wang2020,Ma2020,Hanifi2021,Maitre2021}. The general approach is to train a classifier using extracted time-frequency domain features of reflected Doppler signal. For example, Jin \textit{et al.}\cite{Jin2019} feed a CNN-based classifier with Short Time Fourier Transform(STFT) time-frequency spectrum. With the development of deep learning, researchers realize that the feature extraction operation can be done within the deep neural network. For instance, Maitre \textit{et al.}\cite{Maitre2021} used an FMCW radar to get users' Doppler signals and directly fed them into a classifier based on CNN-LSTM structure.

Recently, a fall detection work system on 4-D radar point cloud was carried out by Jin \textit{et al.}\cite{Jin2020}. 4-D radar point clouds can spatially inform the task of fall detection, containing a wealth of information on human posture, obstacles, etc., and are more robust than previous methods. Jin \textit{et al.} introduced a semi-supervised approach with a Hybrid Variational RNN AutoEncoder (HVRAE). 

\section{Motivation}

The global ageing population puts a greater strain on public healthcare resources due to the declined manpower and more elderly in need of care. Advanced information and sensing technologies are introduced into the healthcare area for daily healthcare monitoring of urgent events \cite{Amin2016}. Among them, falling is a very dangerous signal that needs special attention. It is reported that, even without direct injuries, half of those elders who experience an extended period (more than an hour) of lying on the floor may die within six months after the incident\cite{Amin2016}. So, fall detection is a vital portion of smart healthcare.

\subsection{Limitations of SoA}

\begin{figure}
  \centering
  \includegraphics[width=1\linewidth]{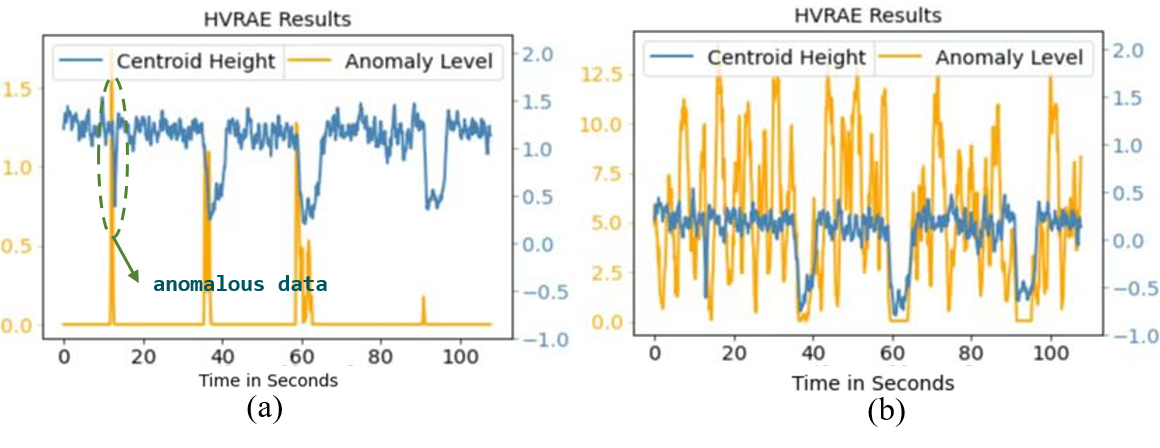}
  \caption{The effect of errors in radar displacements. In each figure, the blue line represents the body's centroid height, and the orange line represents the anomaly level (likelihood of a fall event metric). }
  \label{fig:biasFail}
  \vspace{-3mm}
\end{figure}

Radar-based fall detection is emerging for its unique advantages of privacy-preserving and device-free features over the vision-based and wearable-based approaches. Some state-of-the-art research is based on either the micro-Doppler signal \cite{Tomii2012,Karsmakers2012,Wu2013,Su2015,Sadreazami2019,Jin2019,Takabatake2019,Wang2020,Ma2020,Hanifi2021,Maitre2021,Tian2018}, or the azimuth-elevation-range heatmap \cite{Tian2018}, or the radar point cloud\cite{Jin2020}. These approaches, though work well in the controlled lab environment, suffer from either a high false alarm rate, need considerable training data when being applied in practical scenarios or extremely sensitive to changes in data due to different radar operating environments. What they all have in common is that they use \textbf{environment-dependent features} to train the classifier. For example, an error in radar displacement or some anomalous data from the radar can cause a dramatic drop in system performance. Figure~\ref{fig:biasFail} shows the effect of errors in radar displacement on the performance of mmFall\cite{Jin2020}. Figure~\ref{fig:biasFail}(a) shows the normal output of the mmFall, in Figure~\ref{fig:biasFail}(b) we add -1m to the z-axis data of the test data to simulate the mounting error of the radar in different scenarios. At the same time we find that the anomaly level is also elevated due to some anomalous data from the radar, as shown in the dashed box in Figure~\ref{fig:biasFail}(a).

\subsection{Opportunity}

From the kinematics' perspective, falling is not a monolithic process. A falling process can be further separated into different phases with characteristic features in velocity and acceleration \cite{Noury2016,Seketa2021}. These environment-independent features can be leveraged for high-accuracy and robust fall detection. Though such a motion model is no stranger to those wearable-based fall detection approaches,  it is yet applied in radar-based fall detection systems due to the intrinsic difficulty in extracting clean motion information from the radar signals, such as the velocity and acceleration of the human body, which are environment-independent features and not sensitive to changes in data due to different radar operating environments. Though the basic idea is direct, there are still some real-world challenges that need to be addressed for robust fall detection. These challenges will be detailed in the next subsection.

% Inspired by the stages of a falling process \cite{Noury2016}, we propose a more accurate fall detection framework by integrating the motion models in the radar-based fall detection. Specifically, instead of using all the received signals for fall detection, we focus on the most characteristic stage of a falling process. Such a manner enables the detection of subtle motion of the human body which is essential for differentiating a fall from daily activities.  

\subsection{Challenges}

It is non-trivial to build an accurate and robust fall detection system in practice under a series of practical challenges: 

% \textit{Challenge I: Similar daily activities.} There are some daily activities showing very similar patterns to a falling activity, such as sitting and kneeling. A practical fall detection should be able to capture the subtle differences between them. 

\textit{Challenge I: Environment-independent features estimation.} The actual motion model of a person in a fall is complex and poses a challenge for the extraction of features such as velocity and acceleration. Meanwhile, there are some daily activities showing very similar patterns to a falling activity, such as sitting and kneeling. A practical fall detection should be able to capture the subtle differences between them. 

\textit{Challenge II: Signal scattering.} The presence of scatterers in the environment such as the interior walls and furniture will intervene in the propagation of radar signals by reflecting the signals in all directions. This will create noises such as clutter and ghost targets in the obtained radar image. 

\textit{Challenge III: Multiple users.} A radar system often suffers from a higher false alarm rate when there are multiple users in the radar field of view. That is because the point cloud from different users might be mixed up to downgrade the performance of fall detection. Note that such a problem cannot be well addressed as those noise signals are reflected off static objects like walls and furniture.

Among them, Challenge I mainly affects the accuracy of the system, and Challenge II and Challenge III mainly affect the robustness of the system.

\begin{figure*}
    \centering
      \includegraphics[scale=.55]{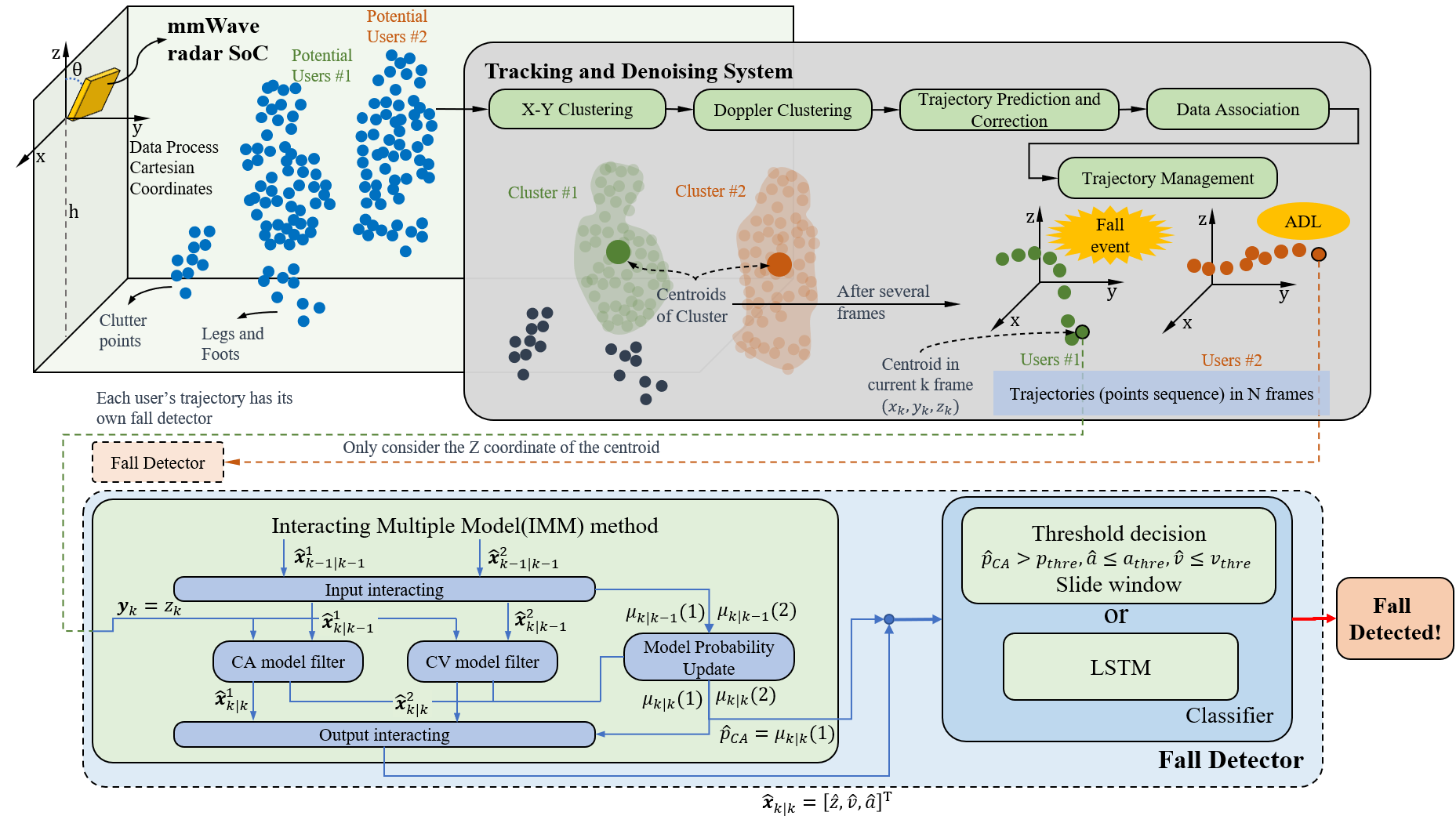}
    \caption{Overview of the workflow in FADE. $h$ is the mounted height of the radar sensor, $\theta$ is the tilt angle of the radar sensor. The mmWave radar generates point cloud data. Tracking and Denoising System uses point cloud data and separates centroids ($x_k,y_k,z_k$) of different users. The input measurement of the IMM module is $\mathbf{y}_k=z_k$. The output of the IMM module is the estimated probability of the CA model (FALL) $p_{CA} = \mu_{k|k}(1)$ and the estimation of user state $\hat{\mathbf{x}}_{k|k}$. In the threshold decision module, $p_{thre}$ is the threshold of CA model, $a_{thre}$ is the threshold of acceleration, $v_{thre}$ is the threshold of velocity.}
    \label{fig:SystemOverview}
\end{figure*}
\section {System Overview}
  
FADE is a fall detection system that exploits the unique properties of mmWave radar. It operates by transmitting an RF signal and recording its reflections off objects. By analyzing the point cloud generated from the reflected signal, it then infers the user's trajectories and detects the existence of fall. The FADE system consists of three modules that operate in a pipeline fashion, as shown in Figure~\ref{fig:SystemOverview}:
\begin{enumerate}
\item mmWave Radar SoC: In this module, an Linear Frequency Modulated Continuous Wave (LFMCW) radar transmits millimeter waves and records the reflections from objects. It then undertakes the radar signal processing and generates the point cloud data.
\item Tracking and Denoising System: This module is built to make the system obtain high robustness which means two things: 1) reducing the impact of clutter and ghost target on the system, 2) separating the information of multiple users that may exist in the scene. The tracking and denoising system intakes the point cloud data from mmWave radar SoC and generates stable sequences of centroids of different potential users. The system robustness design will be introduced in Section~\ref{subsec:Clustering} and Section~\ref{subsec:TrackingTrajectoryManagement}. 
\item Fall Detector: In this module, an IMM algorithm is implemented to increase the accuracy of FADE by estimating the state of users with the sequences of centroids from the tracking and denoising system. A classifier will trigger a fall instance once it's detected and generate an alarm. The system accuracy design will be introduced in Section~\ref{sec:fall_detection}
\end{enumerate}

\section{Accuracy Design: IMM-based Fall Detection} \label{sec:fall_detection}

In this section, we first review the process of falling and summarize the key information that can be used to build FADE. After that, the IMM-based fall detector design is proposed.

\begin{figure}
  \centering
  \includegraphics[width=0.98\linewidth]{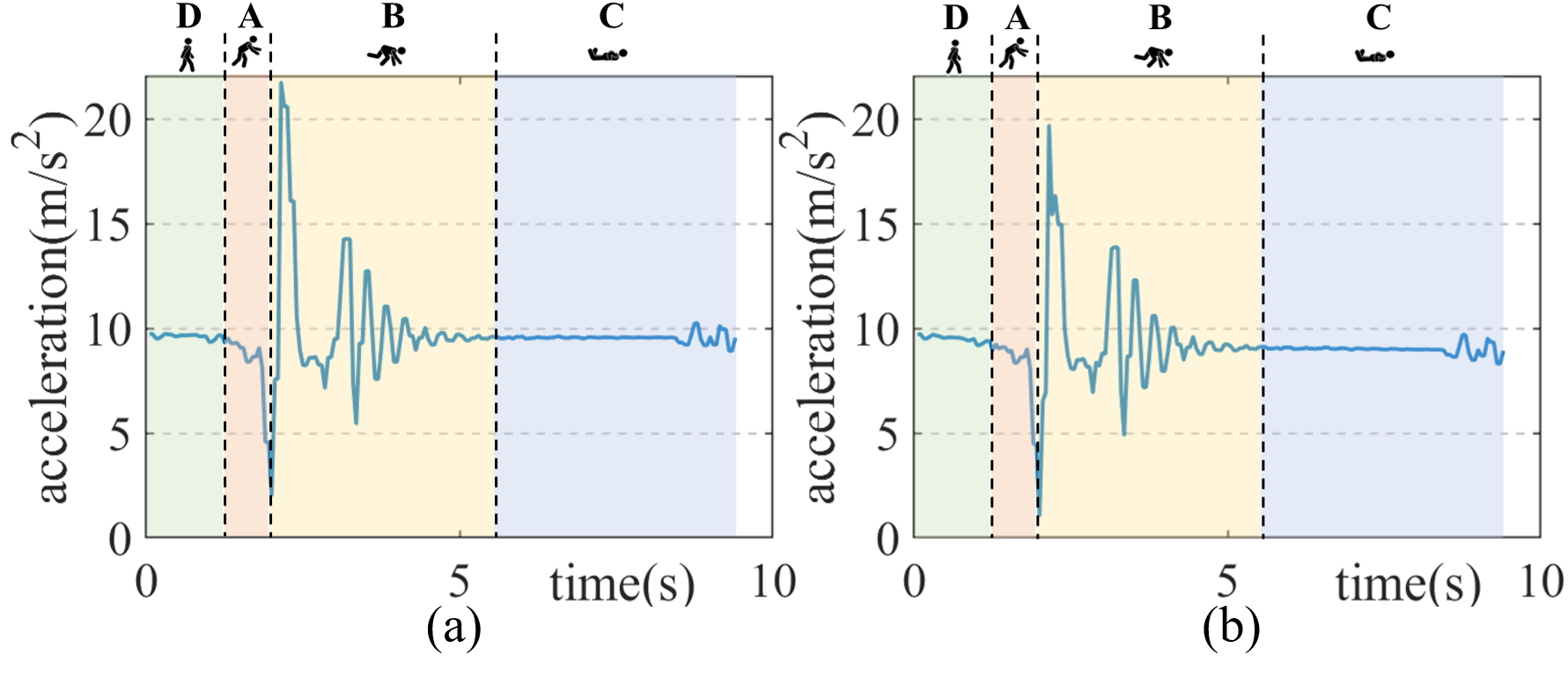}
  \caption{An example of a falling signal from IMU with four phases: The shown data is from UP-Fall Detection Dataset\cite{Martinez-Villasenor2019}. Phase D is ADL, phase A is the pre-fall phase and falling phase, phase B is the impact phase, and phase C is the rest phase. (a) shows the change of amplitude of acceleration vector from the triaxial accelerometer. (b) shows the value obtained by projecting the user's acceleration vector perpendicular to the ground. The attitude angle of IMU was obtained by complementary filtering of angular velocity and acceleration.}
  \label{fig:FallPhase}
  \vspace{-3mm}
\end{figure}

\subsection{Process of Falling}
\label{subsec:ProcessofFalling}
Although falls are diverse in etiologies (causes), circumstances, characteristics, and clinical consequences, the goal of FADE is to correctly detect a fall event. So, knowing the process of a fall helps us understand the fall detection problem better. In kinematics analysis, falls are usually described as a sequence of multiple phases and different phase has their corresponding feature (e.g. height, velocity, acceleration)\cite{Noury2016,Seketa2021}. Noury \textit{et al.}\cite{Noury2016} split the fall scenario into 4 successive phases (\textbf{pre-fall}, \textbf{fall}, \textbf{impact} and \textbf{rest}) and get fall velocity determined from video analysis. Furthermore, Seketa \textit{et al.}\cite{Seketa2021} describe the acceleration changes during the different phases of a fall. The previous research of Noury and Seketa facilitate the design of FADE since the user's velocity and acceleration are the key variables that FADE estimates from point cloud data. The four stages of falling are as follows: 
\begin{enumerate}
\item \textbf{The pre-fall phase} is the time when a person loses balance and begins to descend uncontrollably to the ground. In this stage, there is only some slight change of the person's state is as shown together with \textbf{the fall} phase as phase A of Figure~\ref{fig:FallPhase}(a).
\item \textbf{The fall phase} is the period between the start of the fall and the body impact on a lower surface. During this phase, the body falls due to gravity, and the vertical velocity increases. And the acceleration towards the ground is in most cases much less than $9.81m/s^2(1 g)$. Shown together with the pre-fall phase as phase A of Figure~\ref{fig:FallPhase}(a). And a significant drop on the centroid can be observed by mmWave radar, shown in Figure~\ref{fig:IMM_position_estimation}.
\item \textbf{The impact phase} is the moment when a person hits the ground or some other lower surface for the first time. The impact usually causes an abrupt change of the acceleration direction and the magnitude of acceleration (as a peak of acceleration in phase B of Figure~\ref{fig:FallPhase}(a)). This change is widely used in IMU-based fall detection systems but it is difficult to be observed by radar due to the frame rate and observation error of the millimeter-wave radar point cloud.
\item \textbf{The rest phase} is the phase that the person is lying or sitting on the ground or other lower surface. If the person is unable to move due to the fall, no significant changes in acceleration magnitude can be observed by IMU in this phase\cite{Seketa2021}. And the height of the user's centroid observed by radar remains at a low level.
\end{enumerate}

In the above four stages, abrupt change in the height of the user's centroid (estimated by radar) can be observed in the fall phase. This phase offers us the key feature that can be used to distinguish ADL and FALL. Our fall detector design mainly leverages the prior knowledge that comes from this phase of fall, which is the change in vertical velocity and vertical acceleration. We can find that both features are environment-independent. 
In other words, we choose the Z-axis data from point cloud (as shown in Figure~\ref{fig:SystemOverview}) estimation as the input of the fall detector since Z-axis data carries useful information to estimate the torso level and the relevant motion model. Since only one-dimensional data is used, the computational cost is reduced. 

\subsection{Feature Extraction: Interacting Multiple Model Algorithm in Fall Detection}
The measured Z-axis data of fall instance is shown in Figure~\ref{fig:IMM_position_estimation}. In order to take advantage of the prior knowledge gained through the process of falling, we need to estimate the Z-axis position, vertical velocity, and vertical acceleration of the user’s centroid through Z-axis data. But in the process of falling, the motion pattern of the user's centroid is switched between the motion models of ADL and FALL. FALL often has a great acceleration towards the ground, while the velocity and acceleration of ADL only fluctuate in a relatively small range. This phenomenon causes a model mismatch problem resulting in higher estimation error for those approaches using only a single model, such as the Kalman filter.

To overcome the above problem, we introduce the Interacting Multiple Model (IMM) algorithm to estimate the user's state. The IMM method assumes that the user dynamic characteristics are contained in the following model sets\cite{BarShalom1989}: 
\begin{equation}
  \mathbf{x}_{k}=\mathbf{f}_{r_{k}}\left(\mathbf{x}_{k-1}, \mathbf{u}_{k}\right)+\mathbf{v}_{r_{k}}, \quad r_{k} \in\{1,2, \cdots, d\}
\end{equation}
where $\mathbf{x}_k$ is the user state at time $k$, $r_{k}$ is a random variable that satisfies the consistent discrete Markov chain in the state space $\{1,2, \cdots, d\}$ at time $k$, $\mathbf{v}_{r}$ is the process noise of the model $r$, $\mathbf{f}_{r}(\cdot)$ is the transfer equation of the model $r$, $\mathbf{u}_{k}$ is the unknown system input vector at time $k$.

In order to adapt the IMM framework to the problem of fall detection, first, we model the vertical movement of the user in the FALL and ADL states as constant velocity (CV) model ($r=1$) and constant acceleration (CA) model ($r=2$), respectively. The state vector of the user is $\mathbf{x}=\left[z\ v\ a\right]^T$ and the transfer equation of these two models are 
\begin{equation}
  \mathbf{f}_{1} (\mathbf{x},\mathbf{u})= \mathbf{F}_{1}\cdot \mathbf{x}+\mathbf{u}
  \label{equ:TransferEquation_CV}
\end{equation}
\begin{equation}
  \mathbf{f}_{2} (\mathbf{x},\mathbf{u})= \mathbf{F}_{2}\cdot \mathbf{x}+\mathbf{u}
  \label{equ:TransferEquation_CA}
\end{equation}
where $\mathbf{F}_{1}=\left[\begin{matrix}1&T&0\\0&1&0\\0&0&0\end{matrix}\right]\mathbf{F}_{2}=\left[\begin{matrix}1&T&T^2/2\\0&1&T\\0&0&1\end{matrix}\right]$, $T$ is sampling interval, $\mathbf{u}$ is the system input vector. When it is detected that the motion model is transferred from the CV model to the CA model, the least square method is used to estimate $\mathbf{u}$, and for the rest of the time $\mathbf{u}=[0\ 0\ 0]^T$.

Second, we assume that the switch between ADL and FALL is a Markov process. And the state switching matrix ${\Gamma}$ is
\begin{equation}
    {\Gamma} = \left[
    \begin{matrix}
    Pr( r_k = 1 | r_{k-1} = 1) & Pr( r_k = 2 | r_{k-1} = 1)\\
    Pr( r_k = 2 | r_{k-1} = 1) & Pr( r_k = 2 | r_{k-1} = 2)
    \end{matrix}
    \right]
\end{equation}
where $Pr( r_k = i | r_{k-1} = j)$ model transition probability (e.g. $Pr( r_k = 2 | r_{k-1} = 1)$ indicates the probability that the user will transfer from the ADL model to the FALL model from time $k-1$ to time $k$). 

Thirdly, we establish the measurement model
\begin{equation}
  \mathbf{y}_k = \mathbf{H}\mathbf{x}_k + \mathbf{w}_k
\end{equation}
where $\mathbf{H} = [1\ 0\ 0]$ is the measurement matrix since only the position data of user's state can be measured, $\mathbf{w}_k$ is the noise of measurement, $\mathbf{x}=\left[z\ v\ a\right]^T$ is the user's state in time $k$.

After that, through the iterative calculation of IMM algorithm with the presence of our predefined $\mathbf{F}_{1}$, $\mathbf{F}_{2}$, $\mathbf{v}_{r_k}$, $\mathbf{H}$, $\mathbf{w}_k$, ${\Gamma}$ and the measured $\mathbf{y}_k$ at time $k$, we will get the estimation of user’s state vector ($\hat{\mathbf{x}}_{k|k}=\left[\hat{z}\ \hat{v}\ \hat{a}\right]^T$) and the model posterior probability of ADL ($\mu_{k \mid k}(1)$) and FALL ($\mu_{k \mid k}(2)$). IMM algorithm 
has four key steps: 1) input interacting, 2) model filter, 3) model probability update, 4) output interacting\cite{Challa2011}. And through the process of iterative calculation of these four steps we can get $\hat{\mathbf{x}}_{k|k}$ and $\mu_{k \mid k}(i)$. %More details of IMM algorithm can be found in Appendix.

Finally, we design a fall detector using the estimation $\hat{\mathbf{x}}_{k|k}$ and $\mu_{k \mid k}(i)$ from IMM algorithm. 

\subsection{Fall Detector Design}

\begin{figure}
    \subfigure[IMM position estimation]{
      \includegraphics[width=0.96\linewidth]{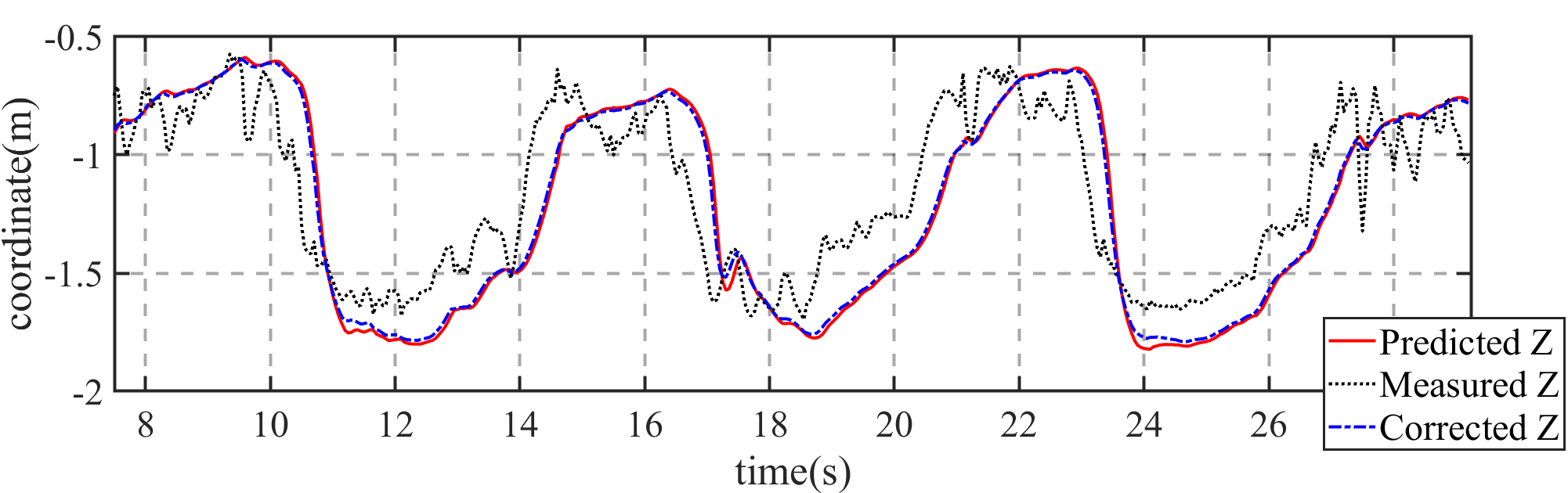}
      \label{fig:IMM_position_estimation}
    }
    \subfigure[IMM velocity estimation]{
      \includegraphics[width=0.9\linewidth]{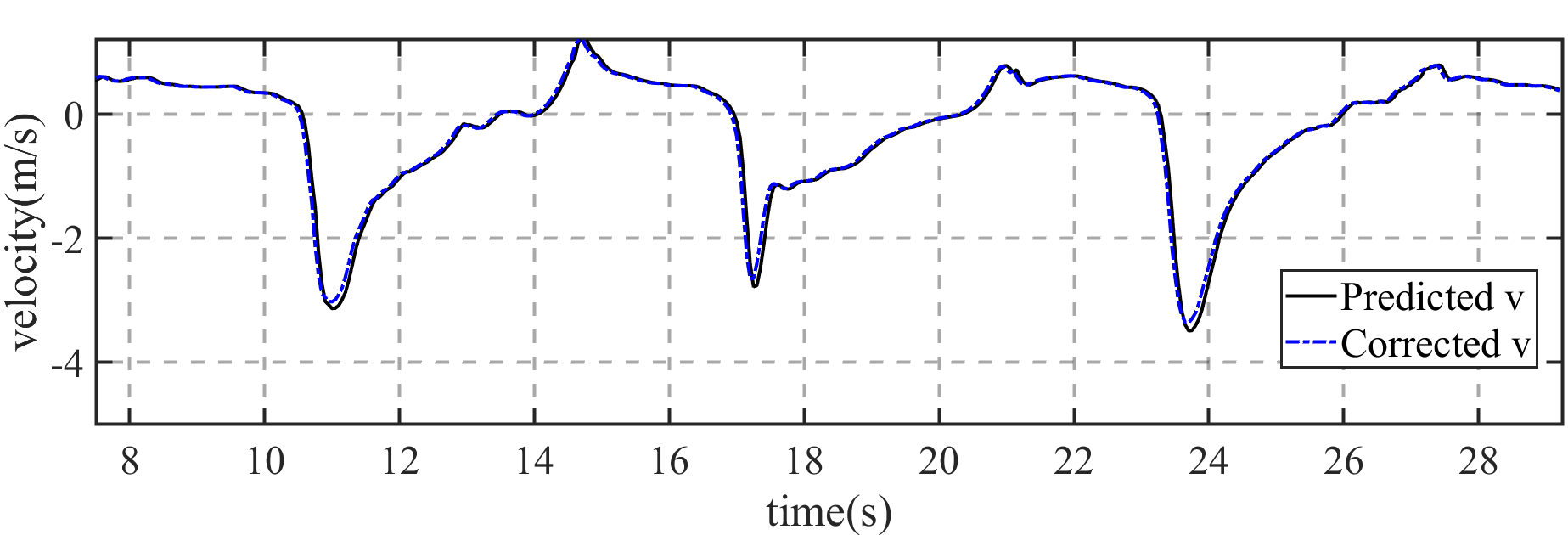}
      \label{fig:IMM_velocity_estimation}
    }
    \subfigure[IMM acceleration estimation]{
      \includegraphics[width=0.9\linewidth]{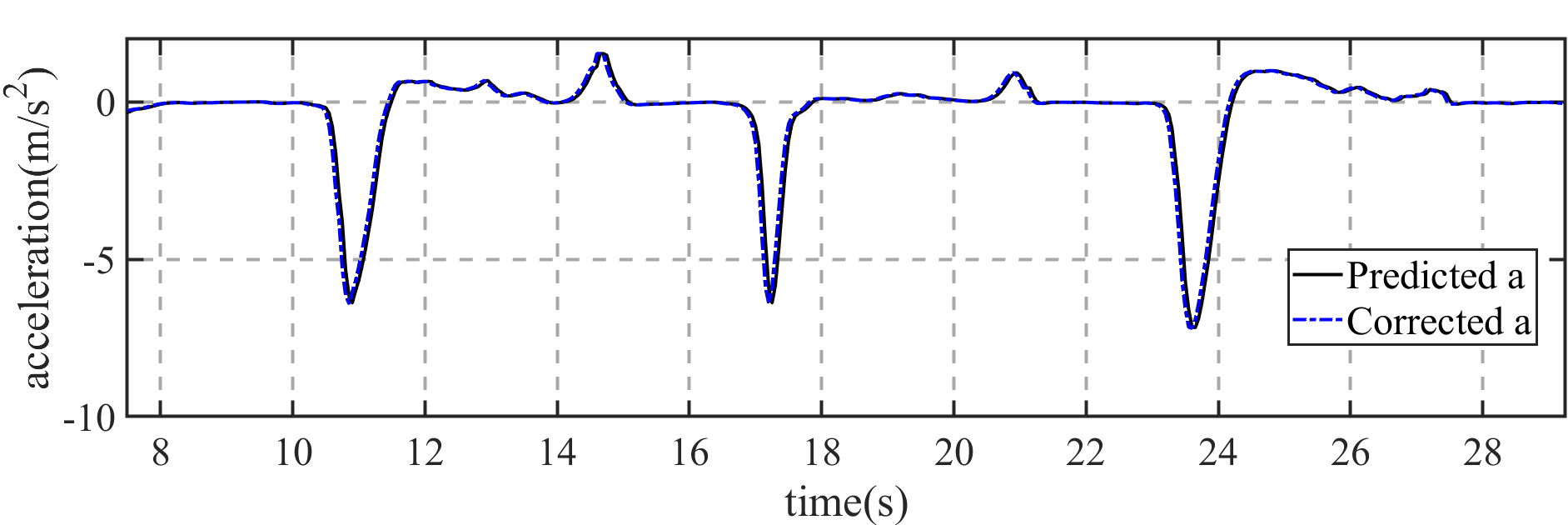}
      \label{fig:IMM_acceleration_estimation}
    }
    \subfigure[IMM model probability estimation]{
      \includegraphics[width=0.9\linewidth]{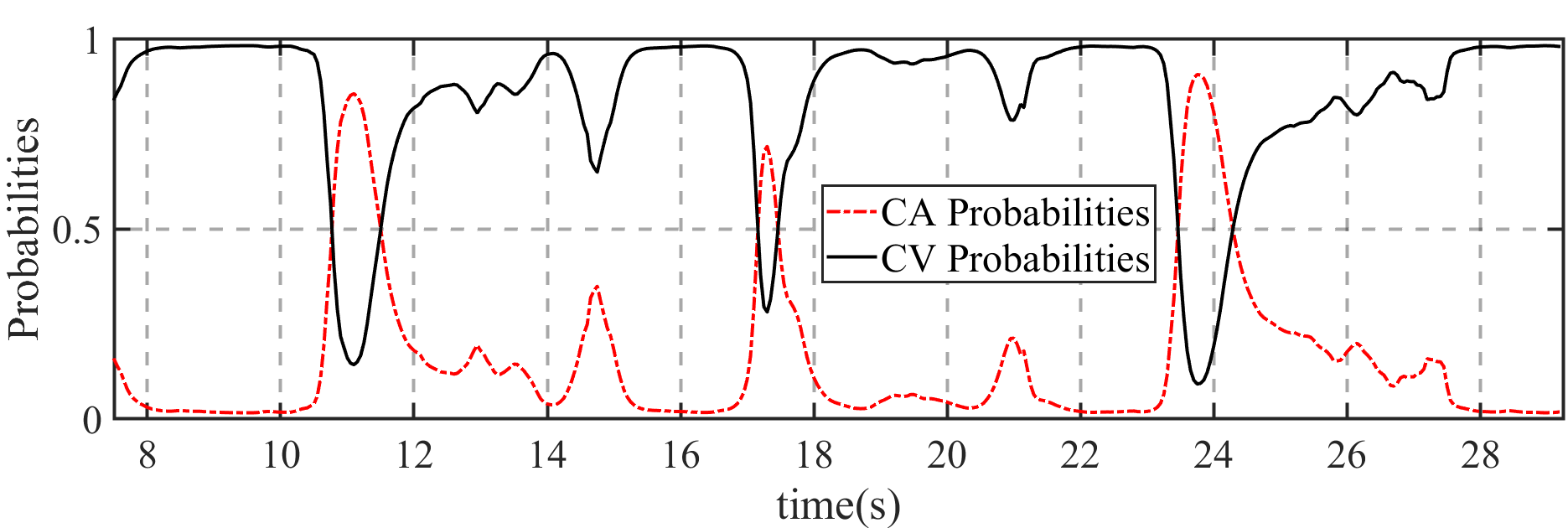}
      \label{fig:IMM_model_probability_estimation}
    }
    \subfigure[Threshold decision]{
      \includegraphics[width=0.9\linewidth]{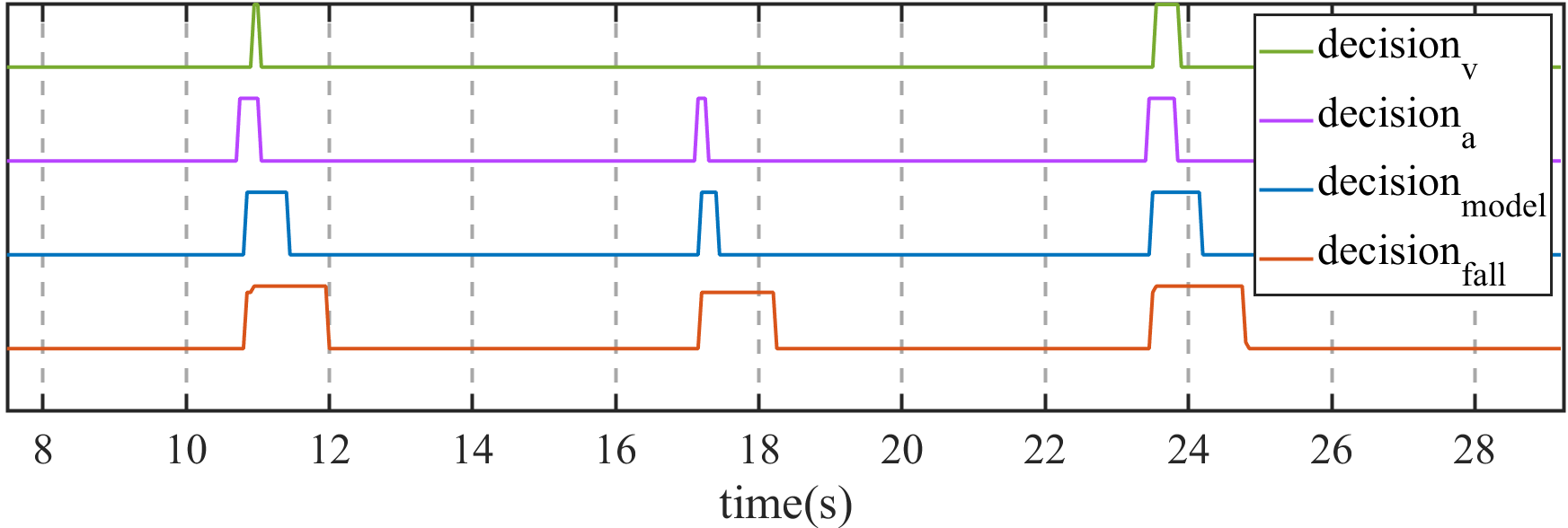}
      \label{fig:threshold_decision}
    }
    \caption{The results of IMM algorithm applied on three real fall instance data: (a) shows the measured, predicted, and corrected Z-axis coordinates of the user's centroid; (b) shows the predicted and corrected vertical velocity data of user's centroid; (c) shows the predicted and corrected vertical acceleration data of user's centroid; (d) shows the posterior model probability; (e) shows the lines of threshold decision output.}
    \label{fig:IMM}
    \vspace{-5mm}
\end{figure}

Considering the complexity of the algorithm implementation, we design two classification methods to implement the fall detector: long short-term memory (LSTM), threshold decision strategy. Figure~\ref{fig:IMM_velocity_estimation},\ref{fig:IMM_acceleration_estimation},\ref{fig:IMM_model_probability_estimation} visualizes the IMM feature (velocity, acceleration, and model probability) extracted from the change of the centrid.

For LSTM method, we train a LSTM based on these three features, a widely used sequential data classifier, to differentiate between fall and ADL.

For threshold decision strategy, we leverage the following two important research data results: (a) Noury \textit{et al.}\cite{Noury2016} revealed the vertical velocity (usually $\mathbf{3m/s}$) during the fall phase. (b) The typical statistical acceleration value from UP-Fall dataset\cite{Martinez-Villasenor2019} is less than $\mathbf{-5m/s^2}$, when a fall instance occurs. Through the threshold decider we get three independent decision lines for velocity, acceleration, and model probability.It is worth noting that the movement of standing up after a fall will also cause a change in acceleration, which increases the probability of FALL model (as shown in Figure~\ref{fig:IMM_model_probability_estimation}). But this change can be easily reduced by the sign of acceleration. After the threshold decision, we use a sliding window detection strategy since the peak of each feature is not arrived simultaneously (as in Figure~\ref{fig:threshold_decision}).

\section{Robustness Design: Tracking and Denoising} \label{sec:system_design}
In this section, we present the robustness design for FADE. This section describes how we process the point cloud data from the radar sensors to obtain reliable, user-specific centriod data that can be delivered to the fall detector. First, we introduce the design of a modified clustering method. Next, we introduce the design of  tracking and trajectory management. 

\subsection{Clustering-Torso Extraction}\label{subsec:Clustering} 
In this section, we propose a grid-based DBscan-like clustering method. This method can 1) find potential user point cloud clusters, 2) separate the torso part of the user’s point cloud, 3) filter out some significant outliers, and 4) reduce the complexity of the calculation.

After the radar signal preprocessing chain, we get a set of point cloud. Each point cloud contains coordinates, signal-to-noise ratio (SNR), and Doppler information, as shown in the Figure~\ref{fig:SystemOverview}. The echo signal of the limbs are not stable signals, which are often accompanied by flickering characteristics. Therefore, the point cloud data processed by the previous signal processing chain is both unstable and uneven, which will cause large fluctuations in the estimation of the centroid. Meanwhile, we focus on the state of the torso when a person falls, so we need to extract the point cloud of the torso during the process of clustering. 

\begin{figure}
  \subfigure[DBscan result]{
		\includegraphics[width=0.46\linewidth]{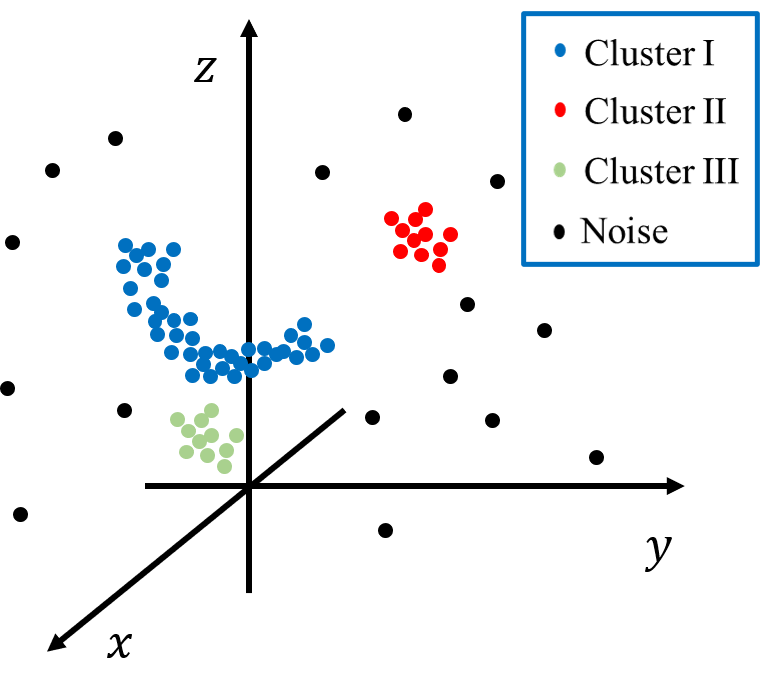}
  }
  \subfigure[Modified DBscan result]{
		\includegraphics[width=0.46\linewidth]{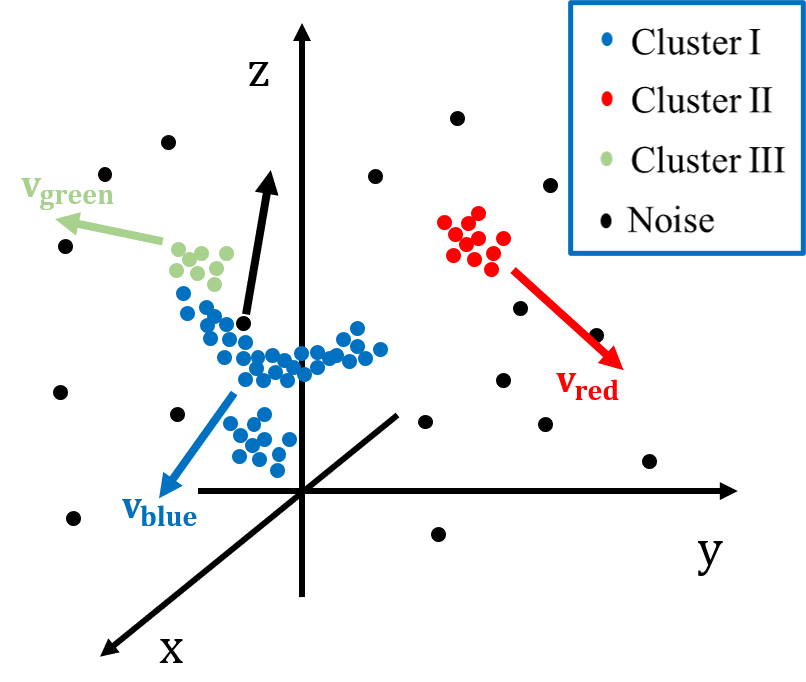}
  }
  \caption{Comparison of point cloud clustering by DBscan and modified DBscan. In (a), point cloud (Clusters I and Clusters III) from the same user are clustered incorrectly due to the large Z-direction distance. In (b), point cloud (Clusters I and Clusters III) are well clustered because of the different Doppler velocities.}
  \label{fig:DBscan}
  \vspace{-3mm}
\end{figure}

Because the torso has a large surface and is easier to scatter electromagnetic waves than the limbs, the points from the body part are uniform and dense (as shown in Figure~\ref{fig:SystemOverview}), and the Doppler velocities are relatively similar. Thus, we change the similarity function of DBscan into 
\begin{equation}
  \begin{aligned}
    D(p^i, p^j)^2 = (p^i_x - p^j_x)^2 + (p^i_y - p^j_y)^2 +
    \\  \alpha(p^i_z - p^j_z)^2+\beta(p^i_v - p^j_v)^2
  \end{aligned}
  \label{equ:similarity_function}
\end{equation}
The performance of such a relatively simple method is satisfying , as shown in the Figure~\ref{fig:DBscan}. 

However, the computational cost of calculating the modified similarity function is not acceptable in our system since it not only required lots of square arithmetic but lots of storage spaces as well. So, We make a direct trade-off between precision, speed, and space. We first let $\alpha = 0$ in Equation \ref{equ:similarity_function}, which means the during the clustering operation we do not consider the z coordinates. And then we encode \textit{x-y} coordinates of each frame to a 2-D grid map, which not only carries coarse position information but also provides an efficient index method of adjacent location. 

We define a 2-dimensional grid $\mathbb{C}$, the $i-th$ cluster $\mathbb{Q}_i$ and the element of the $i-th$ cluster $\mathbb{Q}_i$ is defined as $\mathbf{E}\in \mathbb{C}$, $surround(\mathbf{E})$ as the set of adjacent grids of $\mathbf{E}$, $num(\mathbf{E})$ is the number of point clouds in grid $\mathbf{E}$. The grid-based clustering method is shown in Algorithm~\ref{alg:DBscan}.

\RestyleAlgo{ruled}
\begin{algorithm}[t]
  \label{alg:DBscan}
	\caption{Grid-based DBscan-like clustering}
	\LinesNumbered
	\KwIn{$\mathbb{C}$,$Thre_{Starter}$,$Thre_{Final}$}
  \KwOut{$\mathbf{Q_i}(i=1\cdots n)$}

  Initialize: current cluster index $k \leftarrow 1$, current cluster $\mathbf{Q_i}\leftarrow\phi$, find the biggest gird in $\mathbb{C}$: $\mathbf{E} \leftarrow max(\mathbb{C})$\\
  \While(){$num(\mathbf{E}) \geq Thre_{Starter}$}
  {
    $\mathbf{Q}_k =\leftarrow \mathbf{E}$\\
    $expand(\mathbf{E},\mathbf{Q}_k, \mathbb{C})$ // Regard $\mathbf{E}$ as the center of the grid and group the grids around it into one cluster  \\
    
    \If(){$num(\mathbf{Q}_k)\geq Thre_{Final}$}
    {
      Index increment: $k \leftarrow k+1$
    }
    \text{Update the operating gird} $\mathbf{E} \leftarrow max(\mathbb{C})$ \\
  }
  Return $\mathbf{Q}$
\end{algorithm}

After clustering in \textit{x-y} plane, we get a couple of cluster sets $\mathbb{Q}_i(i = 1,2,\cdots,n)$. To achieve the goal in Equation (\ref{equ:similarity_function}), for every cluster set $\mathbb{Q}_i$, we performed a velocity clustering and choose the set with the biggest velocity to calculate the centroid. 

\subsection{Tracking and Trajectory Management}
\label{subsec:TrackingTrajectoryManagement} 

\begin{figure*}
  \centering
    \includegraphics[scale=.45]{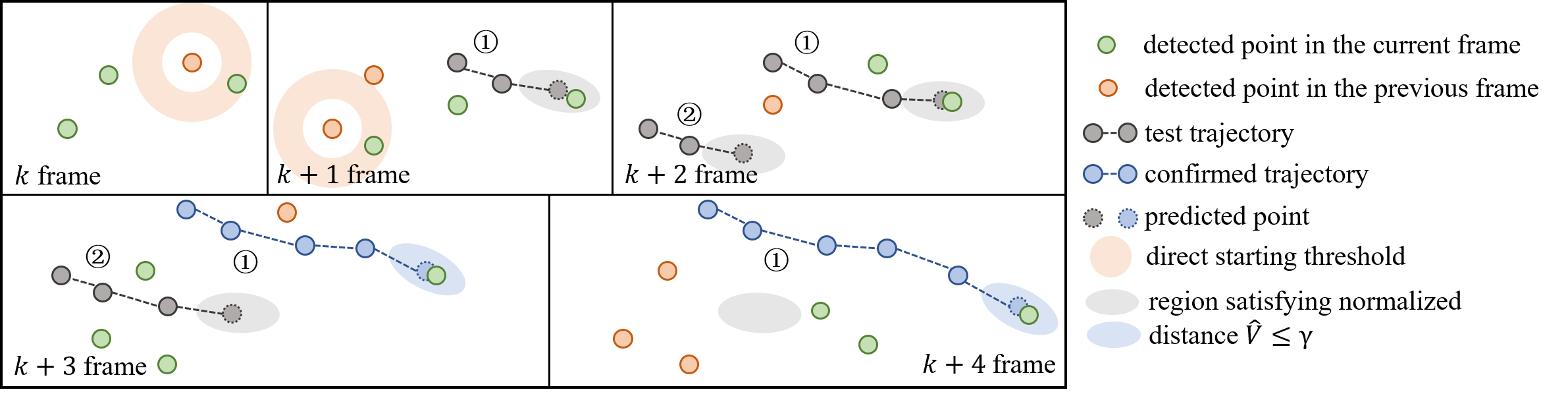}
  \caption{Workflow of tracking and trajectory management. In $k$ frame and $k+1$ shows an example of inter-frame direct starting method. A success direct starting instance is detected in $k$ frame and generate a new test trajectory $\textcircled{1}$ as shown in $k+1$ frame. Then MN-logic starter of test trajectory $\textcircled{1}$ begins to work with $M = 3$, $N = 4$. The detected point satisfying $\hat{V} \leq \gamma$ will be associated and added to test trajectory $\textcircled{1}$ sequence. From $k+2$ frame to $k+3$ frame,  the test trajectory $\textcircled{1}$ successfully changes into a confirmed trajectory by MN-logic method. On the contrary, test trajectory $\textcircled{2}$ is deleted by MN-logic method since it fails to associate any detected points in either $k+2$ or $k+3$ frame.}
  \label{fig:tracking}
\end{figure*}

In the tracking and trajectory management module, we solve the following problems: 1) track multiple users, 2) manage track birth and death, state estimation, prediction and smoothing, data association. This approach guarantees two things: 1) In the fall detection process, we can provide separate services for different users without their data interfering with each other, 2) It can reduce the ghost target interference caused by the multi-path effect to a large extent. Firstly, some simple pairs called test trajectories are obtained through the inter-frame direct starting method. Then, the candidate pairs are judged to be a confirmed trajectory by MN-logic method. For the test trajectories and the confirmed trajectories, we use the probability nearest neighbor (PNN) criterion for data association and use the Kalman Filter (KF) method to update and predict the user's state. Meanwhile, we introduce the scoreboard strategy to delete the confirmed trajectories when the user leaves the scene, for example. The workflow of tracking and trajectory management module is shown in Figure~\ref{fig:tracking}.

\subsubsection{Inter-frame Direct Starting}
The common method of inter-frame direct starting is based on the maximum and minimum speed constraints. Assuming that the measurement set at time $t$ is $\mathbb{Z}_t=\{\mathbf{Z}^t_1, \mathbf{Z}^t_2, \cdots, \mathbf{Z}^t_n\}=\{(x^t_1,y^t_1,z^t_1),(x^t_2,y^t_2,z^t_2),\cdots,(x^t_n,y^t_n,z^t_n)\},n=|\mathbb{Z}_t|$, the measurement set at time $t-1$ is $\mathbb{Z}_{t-1}=\{\mathbf{Z}^{t-1}_1\,\mathbf{Z}^{t-1}_2\,\cdots,\mathbf{Z}^{t-1}_m\}=\{(x^{t-1}_1,y^{t-1}_1,z^{t-1}_1),(x^{t-1}_2,y^{t-1}_2,z^{t-1}_2),\cdots,(x^{t-1}_m,y^{t-1}_m,z^{t-1}_m)\},m=|\mathbb{Z}_t|$, we calculate the distance between two frames:
\begin{equation}
  \begin{aligned}
    D_{ij} = \sqrt{(x^t_i-x^{t-1}_j)^2+(y^t_i-y^{t-1}_j)^2+(z^t_i-z^{t-1}_j)^2},\\(1\leq i \leq n,1\leq j \leq m)
  \end{aligned}
\end{equation}
If it satisfies
\begin{equation}
V_{min}T_s\leq D_{ij}\leq V_{max}T_s
\end{equation}
We put such a pair $\mathbf{T}_k=[(x^{t}_i,y^{t}_i,z^{t}_i),(x^{t-1}_j,y^{t-1}_j,z^{t-1}_j)]$ into the test trajectory set ($\mathbb{T}$), where $V_{min}$ is the minimum speed value of the constraint, $V_{max}$ is the maximum speed value of the constraint and $T_s$ is the interval at which frames are sampled.

\subsubsection{MN-logic Trajectory Starter and Finisher}
The inter-frame direct starting method can only bring some candidate test trajectories, so we need a certain strategy to identify which of the test trajectories are generated by clutter points and which are the possible trajectories of real users. 

We use MN-logic method to deal with this problem. Assuming that the test trajectories $\mathbb{T}$ set has a new element $\mathbf{T}_k=[\mathbf{Z}^{t-N+2}_{i_{t-N+2}},\mathbf{Z}^{t-N+1}_{i_{t-N+1}}]$ at time $t-N+2$, it should be noticed that $\mathbf{T}_k$ is a sequence and we can append other coordinates to it by the data association method. Then at time $t$, $\mathbf{T}_k$ becomes $\mathbf{T}_k=[\mathbf{Z}^{t}_{i_{t}},\mathbf{Z^{t-1}_{i_{t-1}}},\cdots,\mathbf{Z}^{t-N+2}_{i_{t-N+2}}, \mathbf{Z}^{t-N+1}_{i_{t-N+1}}]$ by performing data association method, where
$$
  \mathbf{Z^{t}_{i_{t}}}=\left\{
  \begin{matrix}
  & (x^t_{i_{t}},y^t_{i_{t}},z^t_{i_{t}}) \ & \text{successfully associate with data}\\
  &\phi \ & \text{failed to associate with data}
  \end{matrix}
  \right.
$$
  
At time $t$, the length of $\mathbf{T}_k$ is $N$, Let $M$ be the number of non-empty elements of $\mathbf{T}_k$. When $M$ is not less than a certain threshold set by us (this threshold should not be greater than $N$), we can change the test trajectory $\mathbf{T}_k$ into a confirmed trajectory $\mathbf{C}_k$. If $M$ is less than a certain preset threshold, we consider this test trajectory as a trajectory caused by clutter interference.

Similarly, we apply such MN-Logic method to the management of the confirmed trajectories. We design a scoreboard for each confirmed trajectory to indicate its tracking quality. The scoreboard records the number of times that a confirmation trajectory $\mathbf{C}_k$ fails to associate data in N consecutive frames. When $M$ is less than a certain preset threshold (this threshold should not be greater than $N$), we consider the confirmed trajectory to be missing and delete it from the confirmed trajectory set($\mathbb{C}$).
\subsubsection{Probability Nearest Neighbor Data Association}

Probability nearest neighbor data association method is based on several assumptions\cite{Challa2011}: 1) The measurement noise is White Gaussian noise 2) The motion characteristics of the target follow linear Gaussian statistics. The Mahalanobis distance between the track prediction point and the quantitative meteorological selection point is calculated at each interconnection of the track. The wave gate is set by selecting the threshold value of the normalized distance, and the points falling into the wave gate are interconnected by using the nearest neighbor (NN) criterion.

The normalized distance $V$ is calculated as follows:

If the measured vector (in Cartesian coordinates) $\mathbf{Z}_c(k+1)$ satisfy:
\begin{equation}
  \begin{aligned}
  \hat{V}_{k+1}(\gamma)
  =&\left[\mathbf{Z}_c(k+1)-\hat{\mathbf{Z}}(k+1 \mid k)\right]^{T} \mathbf{S}^{-1}(k+1)\\
  &\left[\mathbf{Z}_c(k+1)-\hat{\mathbf{Z}}(k+1 \mid k)\right] \\
  =&\mathbf{v}^{T}(k+1) \mathbf{S}^{-1}(k+1) \mathbf{v}(k+1) \leq \gamma
  \end{aligned}
\end{equation}
where $\hat{\mathbf{Z}}_{c}(k+1 \mid k)$ is the predicted value of the measurement at time $k$ to time $k+1$, $\mathbf{S}(k+1)$ is the covariance matrix of the measurement prediction at time $k$ to time $k+1$. The measured value $\mathbf{Z}_{c}(k+1)$ is called the candidate measurement, and the above equation is named as the elliptic wave gate rule\cite{Challa2011}. Both $\mathbf{S}(k+1)$ and $\hat{\mathbf{Z}}_{c}(k+1 \mid k)$ are obtained during the prediction process of Kalman filter. The parameter $\gamma$ is obtained from the $\chi^{2}$ distribution table. If the measured value $\mathbf{Z}_{c}(k+1)$ is $n_{{Z}}$ dimensional, then $\hat{V}_{k+1}(\gamma)$ is $\chi^{2}$  distribution random variable with $n_{{Z}}$ degrees of freedom after standardizing the residual. This makes the nearest-neighbor criterion probabilistically optimal under different noise conditions and user velocity conditions.

\section{Implementation and Evaluation} \label{sec:evaluation}
\begin{figure}
  \centering
  \includegraphics[width=0.75\linewidth]{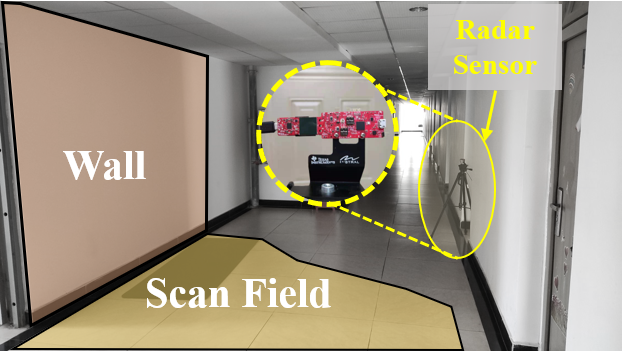}
  \caption{The evaluation environment of FADE.}
  \label{fig:testscene}
  \vspace{-3mm}
\end{figure}
In this section, we present some implementation details of the system and experimental evaluation of FADE. We first conducted the overall performance for FADE: \textit{accuracy}, \textit{robustness} and \textit{time cost} and compare it with other solutions. Then, we conduct the component study to reveal the impact of the tracking and denoising system. After that, we explain the reason that our proposed system is more suitable for feasible fall detection than the Doppler-based method.

\subsection{Implementation and Experiment Set Up}
We deployed FADE (IMM + Thresheld Decision) on the Texas Instrument (TI) IWR6843AOPEVM mmWave FMCW radar evaluation board and place it in the experimental scene (in Figure~\ref{fig:testscene}). We use the C programming language to implement the fall detection algorithm and flash the program into the on-board flash. Besides, the size of the program running on the chip is only 509KB.

\begin{figure}
  \centering
  \includegraphics[width=1\linewidth]{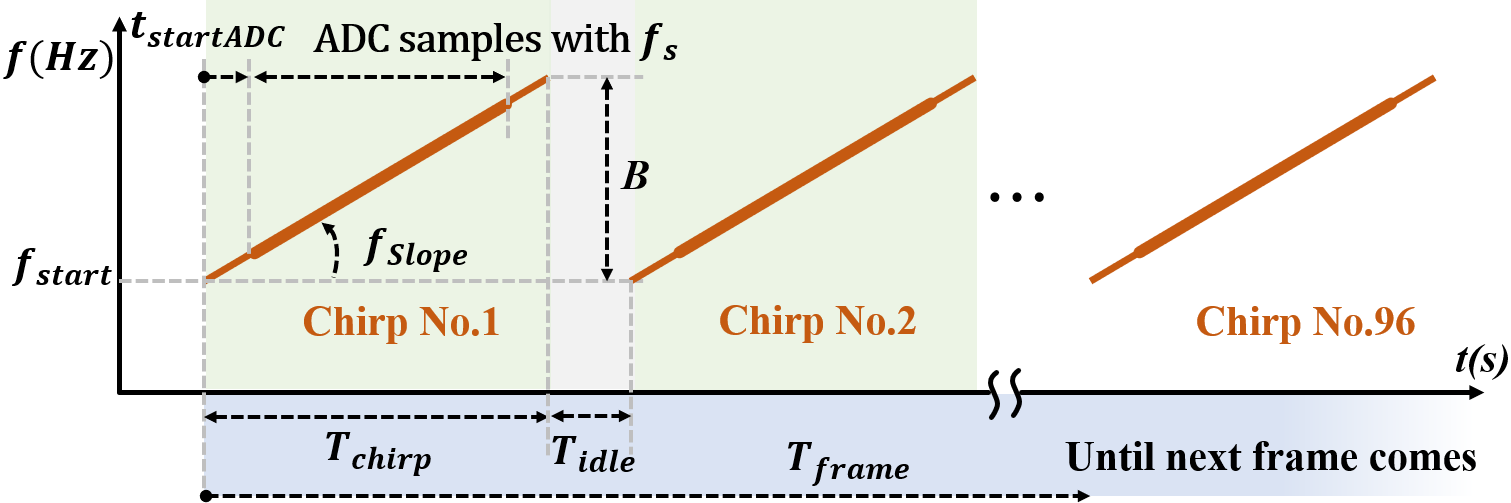}
  \caption{Radar waveform parameters of FADE: The chirp waveform repeats $96$ times in each frame and the period of frame $T_{frame} = 50ms$. $f_{start} = 60.75GHz$ is the start frequency. $f_{slope} = 54.71MHz/ \mu s$ is the frequency slope. $T_{chirp} = 41.1\mu s$ is the time duration of transmitting a chirp signal. $T_{idle}$ is the time interval between two chirps. The ADC start sampling $t_{startADC} = 7\mu s$ after start transmitting the chirp signal and its sampling rate $f_s=2.95MHz$}
  \label{fig:ChirpDiagram}
  \vspace{-3mm}
\end{figure}

The waveform parameters of FADE is shown in Figure~\ref{fig:ChirpDiagram}. With these parameter settings, the working bandwidth of the system is $B = 2.25GHz$, range resolution is $\Delta R= 0.0843m$, speed resolution is $\Delta v = 0.1797m/s$, maximum unambiguous range is $R_{max} = 7.28m$, maximum unambiguous speed is $v_{max} = 8.5556 m/s$.

\subsection{Methodology}

\subsubsection{Data Collection}
During the experiment, we collect point cloud data from three volunteers and construct four data sets as listed in Table~\ref{tab:Dataset}. First, the FALL DS1 data set contains 104 falls in total for fall detector performance evaluation. Second, in FALL DS2 and FALL DS3, we collect data of falling activities from multiple people for evaluation of system performance under the multi-user scenario. Lastly, we constructed the dataset ADL DS1 consisting of Activities of Daily Living.
\begin{table}[]
  \centering
  \caption{Collected Dataset}
  \label{tab:Dataset}
  \begin{tabular}{|c|l|}
  \hline
  Name     & \multicolumn{1}{c|}{Description} \\ \hline
  FALL DS1 & \begin{tabular}[c]{@{}l@{}}1 user in the scene with 50 falls from left to right, \\ 29 falls towards radar,24 falls against radar.\end{tabular}     \\ \hline
  FALL DS2 & 2 users in the scene with 34 falls   \\ \hline
  FALL DS3 & 3 users in the scene with 31 falls \\ \hline
  % FALL DS4 &  falls in different bins \\ \hline
  ADL DS1  & \begin{tabular}[c]{@{}l@{}}45000 frames of ADL, including deep squats, sit down \\ to a chair, etc. And it has 2 falls for validation.\end{tabular} \\ \hline
  \end{tabular}
  \vspace{-3mm}
  \end{table}
\subsubsection{Metrics}
The goal of a fall detection system is to reduce false alarms as well as to avoid missed falls. Thus, we use the following metrics that can express the sensitivity and specificity of fall detection systems. We use True Positives (TP) for the correct detection of falls, False Negatives (FN) for missed falls and False Positives (FP) to indicate false alarms.
\begin{itemize}
\item Precision: The fraction of correctly detected falls over all detected falls, $ Precision = \frac{TP}{TP+FP}$
\item Recall: The fraction of correctly detected falls over the total number of falls $ Recall = \frac{TP}{TP+FN}$
\item F1 score: The harmonic mean of precision and recall $ F1\ score = \frac{2\times Precision\times Recall}{Precision+Recall}$
\end{itemize}

Notice that the false alarms appear randomly on every data set, therefore, the precision is only meaningful when we compare different methods on the same FALL data set.

\subsubsection{Comparison}
To show that our system outperforms the state-of-the-art Radar-based fall detection works in terms of accuracy and robustness to diverse working conditions, we choose mmFall\cite{Jin2020} and a Doppler-based based method as our baseline. We refer to \cite{Jin2019} and train a CNN using STFT spectrum of Doppler signal for fall detection. Meanwhile, the Kalman filter-based method is also compared to show the superiority of IMM method. Finally, we evaluate the system performance of Doppler-based method in a multi-user scene and compare it with FADE.

% (the structure of CNN is shown in the Figure~\ref{fig:CNNstructure})
% \begin{figure}
%   \centering
%   \includegraphics[width=0.85\linewidth]{./figure/CNN_structure.png}
%   \caption{The structure of designed CNN}
%   \label{fig:CNNstructure}
%   \vspace{-3mm}
% \end{figure}

\subsection{Overall Performance}
In this section, we first compare the overall performance of FADE with the state-of-the-art radar-based fall detection methods on system accuracy. Then, we evaluate the performance of FADE in multiplayer scenarios. Lastly, we conduct a detailed evaluation of the time consumption of FADE.

\subsubsection{Accuracy}
To highlight the performance of FADE that achieves a better recall rate and a relatively low false alarm rate, we perform an extensive comparison between various methods. We first reproduce the three methods mentioned in mmFall: Hybrid Variational RNN AutoEncoder (HVRAE), Hybrid Variational RNN AutoEncoder-Simplified LOSS (HVRAE-SL) and RNN AutoEncoder (RAE). These three methods are trained using most of the ADL DS1 data, and the system was finely tuned to achieve the highest F1 score on the FALL dataset. For the Doppler-based approach, we train the CNN using half of the data from FALL DS1 and ADL DS1, and evaluate the system on the remaining fall dataset. Data augmentation methods including scaling and sliding windows are used before training. Further, we extract the features of FALL DS1 and ADL DS1 using Kalman filter and IMM algorithm respectively, and used these features to evaluate the threshold-decision-based and LSTM-based fall detectors respectively. The results are shown in the table \ref{tab:Comparison}, where the combination of IMM as the feature extraction method and LSTM as the classifier achieves the highest F1 Score of 0.9537. We also find that the combination of IMM and threshold decision performs well with an F1 Score of 0.9528.

% \begin{table}[]
%   \centering
%   \label{tab:Comparison}
%   \caption{The Comparison Between Different Methods}
%   \begin{tabular}{cccc}
%   \hline
%   \multicolumn{1}{c}{Method} & Recall & Precision & F1 Score \\ \hline
%   HVRAE                      & 0.8835 & 0.7647    & 0.8198   \\
%   HVRAE\_SL                  & 0.9126 & 0.8826    & 0.8974   \\
%   RAE                        & 0.9126 & 0.8624    & 0.8868   \\
%   CNN                        & 0.9903 & 0.7010    & 0.8209   \\
%   KF                         & 0.9619 & 0.8016    & 0.8745   \\
%   KF+LSTM                    & 0.8824 & \textbf{0.9677}    & 0.9231   \\
%   IMM                        & 0.9619 & 0.9439    & 0.9528   \\ 
%   IMM+LSTM                   & \textbf{1.000} & 0.9115    & \textbf{0.9537}   \\ \hline
%   \end{tabular}
% \end{table}

\begin{table}[]
    \centering
    \label{tab:Comparison}
    \caption{The Comparison Between Different Methods}
    \begin{tabular}{ccccc}
        \hline
        \multicolumn{2}{c}{Method}       & Recall          & Precision       & F1 Score        \\ \hline
        \multicolumn{2}{c}{HVRAE\cite{Jin2020}}        & 0.8835          & 0.7647          & 0.8198          \\
        \multicolumn{2}{c}{HVRAE\_SL\cite{Jin2020}}    & 0.9126          & 0.8826          & 0.8974          \\
        \multicolumn{2}{c}{RAE\cite{Jin2020}}          & 0.9126          & 0.8624          & 0.8868          \\
        \multicolumn{2}{c}{CNN\cite{Jin2019}}          & 0.9903          & 0.7010          & 0.8209          \\
        \multicolumn{2}{c}{KF+Thresheld Decision}           & 0.9619          & 0.8016          & 0.8745          \\
        \multicolumn{2}{c}{KF+LSTM}      & 0.8824          & \textbf{0.9677} & 0.9231          \\ \hline
        \multirow{2}{*}{FADE} & IMM+Thresheld Decision      & 0.9619          & 0.9439          & 0.9528          \\
                              & IMM+LSTM & \textbf{1.0000} & 0.9115          & \textbf{0.9537} \\ \hline
    \end{tabular}
    \vspace{-3mm}
\end{table}

% \begin{figure}
%   \centering
%   \includegraphics[width=0.75\linewidth]{./figure/Accuracy.eps}
%   \caption{System performance comparison between IMM, Kalman filter(KF), HVRAE, HVRAE\_SL and RAE. All these methods are fine tuned on FALL DS1 dataset with the same recall rate.}
%   \label{fig:Accuracy}
% \end{figure}

\begin{figure}
  \centering
  \includegraphics[width=0.95\linewidth]{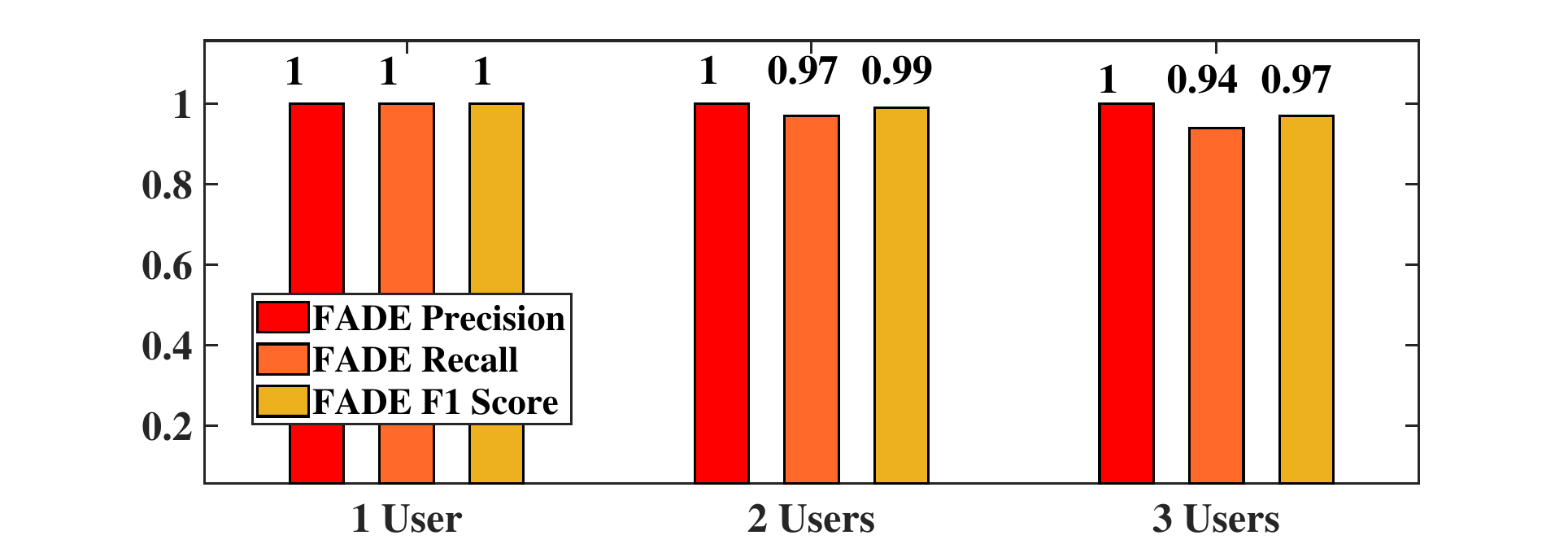}
  \caption{Evaluation of system robustness. The precision, recall rate and F1 score varies with the number of people in the scene.}
  \label{fig:Rubustness}
  \vspace{-3mm}
\end{figure}

\subsubsection{Robustness}
% To make the system work well in multi-user scenarios, it is critical to extract every user's motion information and evaluate the change in the recall rate and precision as the number of users changes. However, the previous work\cite{Jin2020} did not explore related issues. What needs attention in multi-user scenarios is not only the change in recall rate, since such a fall can often be rescued by people around him in time, but also the false alarms caused by interference in multi-user scenarios. Thus, in addition to the accuracy, we would like to understand how robust FADE is in multi-user scenarios, where users would block each other's reflected signals causing point cloud data instability. As shown in Figure~\ref{fig:Rubustness}, FADE consistently detects falls in multi-user scenarios but there is a slight decrease in recall rate with the number of users changed while maintaining a high precision of up to 1.
To make the system work well in multi-user scenarios, it is critical to extract every user's motion information and evaluate the change in the recall rate and precision as the number of users changes. What needs attention in multi-user scenarios is not only the change in recall rate, since such a fall can often be rescued by people around him in time, but also the false alarms caused by interference in multi-user scenarios. Thus, in addition to the accuracy, we would like to understand how robust FADE is in multi-user scenarios, where users would block each other's reflected signals causing point cloud data instability. As shown in Figure~\ref{fig:Rubustness}, FADE consistently detects falls in multi-user scenarios but there is a slight decrease in recall rate with the number of users changed while maintaining a high precision of up to 1.

\subsubsection{Time Cost}
\begin{figure}
    \centering
    \includegraphics[width=0.85\linewidth]{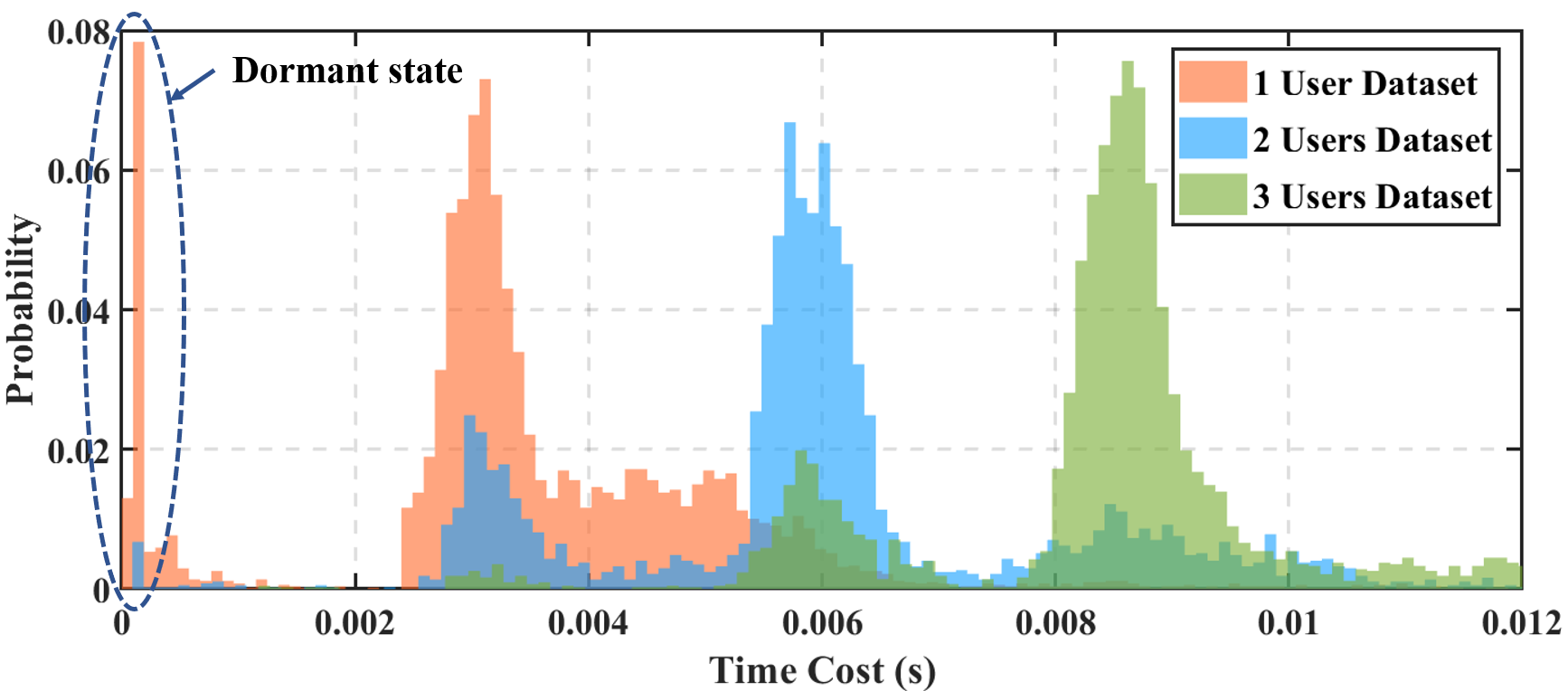}
    \caption{Statistical time cost of FADE on datasets}
    \label{fig:TimeCost}
    \vspace{-3mm}
\end{figure}

\begin{figure}
    \centering
    \includegraphics[width=0.75\linewidth]{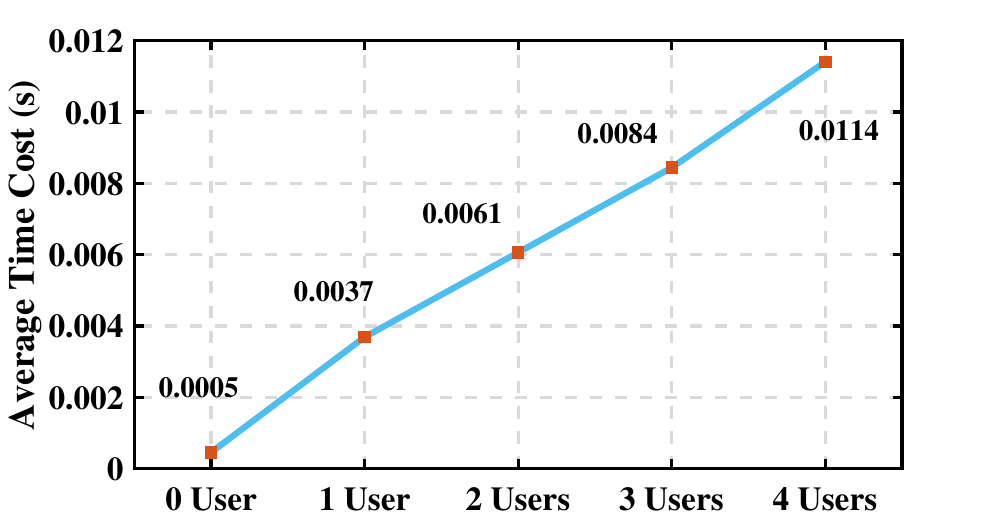}
    \caption{Average time cost of FADE of different number of users}
    \label{fig:TimeCostPer}
    \vspace{-3mm}
\end{figure}

Past works on fall detection are mostly learning-based classifiers driven by a series of neural networks which face the challenge of real-time processing. However, the combination of IMM plus threshold decisions we propose in FADE can be easily be adapt to any indoor radar processor. We record the time cost of FADE (implemented in Matlab) in each frame of FALL DS1, FALL DS2 and FALL DS3 (as shown in Figure~\ref{fig:TimeCost}) to evaluate the change in time cost per frame that varies with the number of people in the scene. In Figure~\ref{fig:TimeCost}, a peak circled on the left side of the picture indicates the dormant state of FADE when there is no user detected in the scene and the system will not be triggered. And each new user detected by the FADE will bring about 3 milliseconds of computational burden to the system, as shown in Figure~\ref{fig:TimeCostPer}.

\subsection{Impact of Tracking and Denoising System}
\begin{figure}
% \vspace{-4mm}
    \subfigure[Before the Tracking and Denoising System]{
          \includegraphics[width=0.45\linewidth]{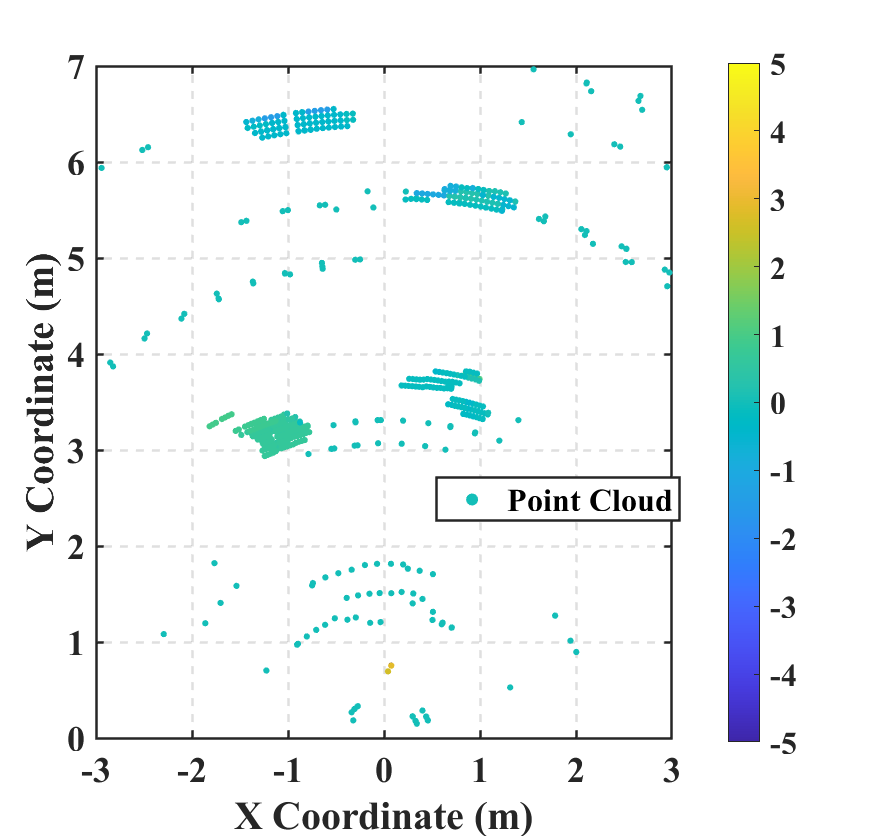}
      \label{fig:IoTDSysBefore}
      }
    \subfigure[After the Tracking and Denoising System]{
          \includegraphics[width=0.45\linewidth]{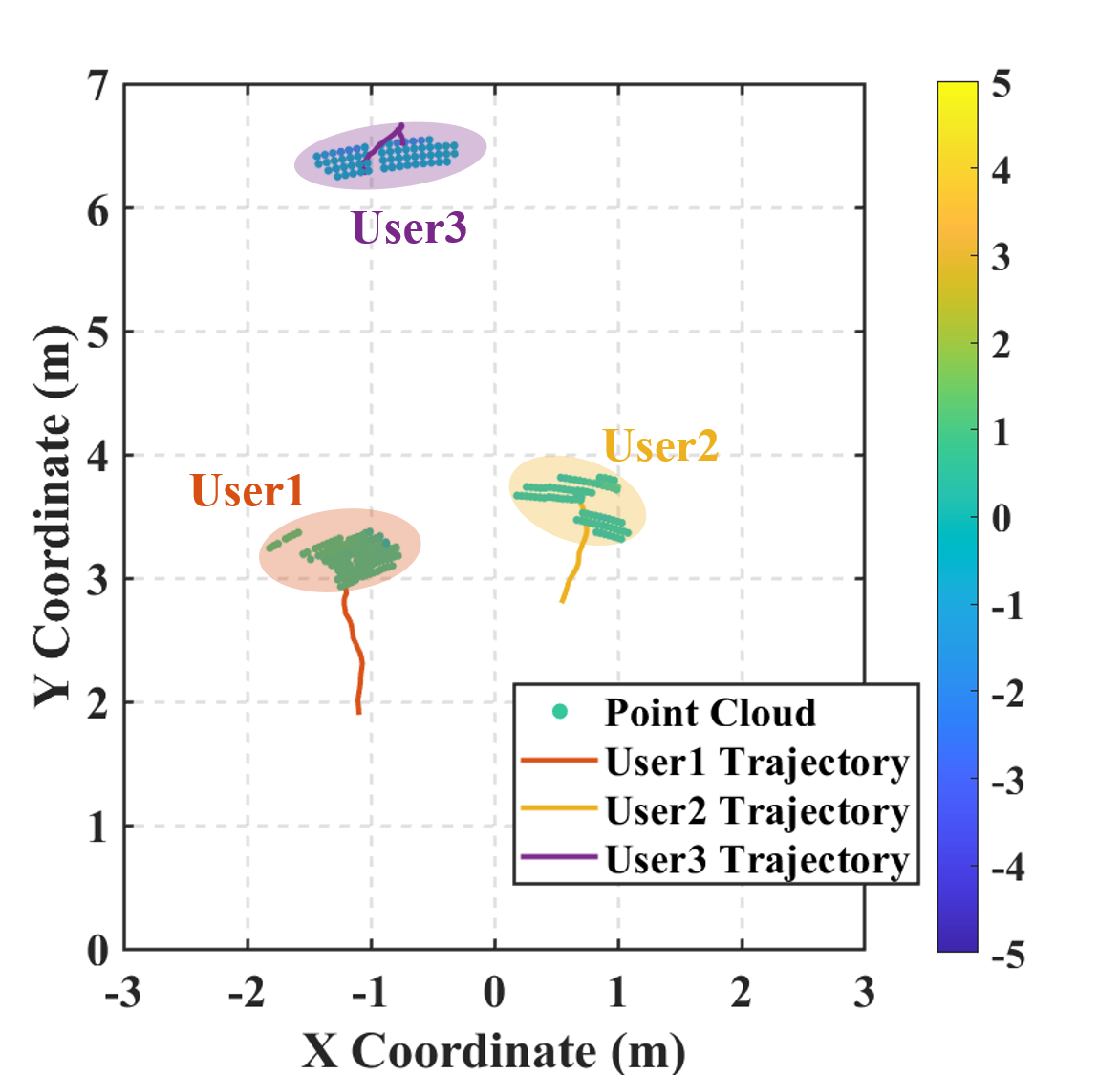}
      \label{fig:IoTDSysAfter}
    }
    \caption{Impact of the tracking and denoising system: the color of the point cloud indicates the velocity of each point and the user trajectory shows the movement of user's centroid in the past 2 seconds.}
    \label{fig:IoTDSys}
\end{figure}

In this section, we examine the impact of the proposed tracking and denoising system. The tracking and denoising system is designed to remove the interference of ghost targets and clutter, as well as to provide each user's information during tracking. Figure~\ref{fig:IoTDSys} shows the input and output of tracking and denoising system. In the upper right corner of the Figure~\ref{fig:IoTDSysBefore}, there is a cluster of points from ghost target caused by reflections from the wall. After the process of tracking and denoising system, not only the clutter is eliminated, but also the ghost target (as shown in Figure~\ref{fig:IoTDSysAfter}).

\begin{figure}
    \centering
    \subfigure{
          \includegraphics[width=0.45\linewidth]{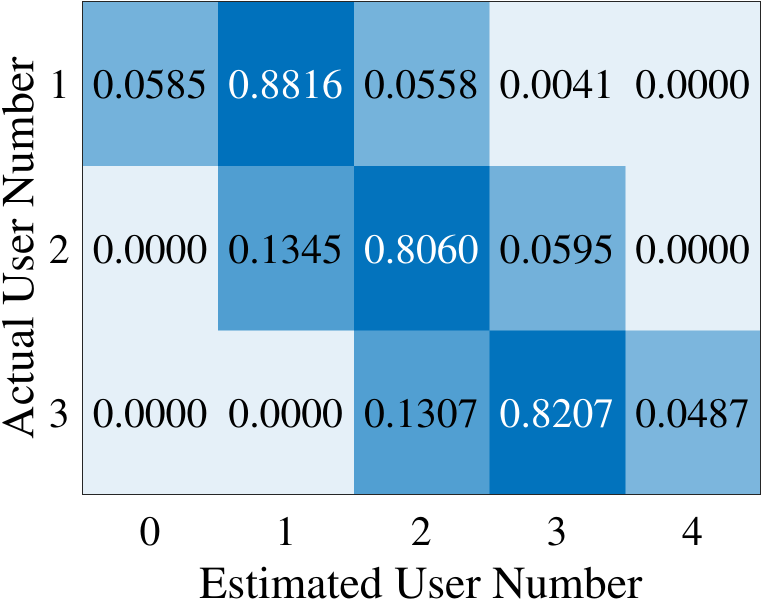}
      \label{fig:usernum}
      }
    % \hspace{-4mm}
    \subfigure{
          \includegraphics[width=0.45\linewidth]{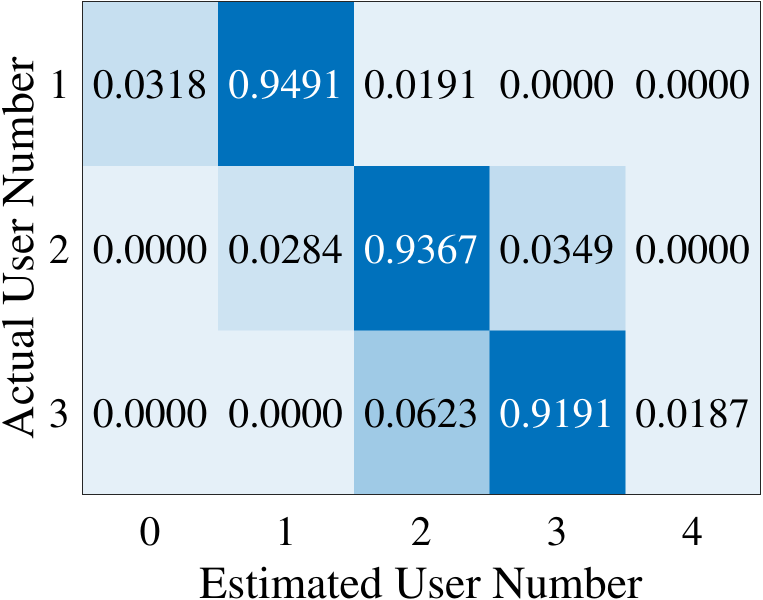}
      \label{fig:usernum_C}
    }
    \caption{(a) Estimation of users' number on FALL DS1, DS2, DS3. (b) Estimation of users' number in a controlled situation (fully active).}
\end{figure}

In order to give a quantitative evaluation, we evaluate the user number estimation performance of the proposed tracking and denoising system (as shown in Figure~\ref{fig:usernum}) on FALL DS1, FALL DS2 and FALL DS3. As can be seen, the system yields an overall accuracy of around $80\%$ in user number estimation. The result is due to the interference between users on the one hand, and the elimination of stationary targets in the radar signal processing on the other hand. On that basis, when the users in the scene are fully active, the user number estimation accuracy in such condition is shown in Figure~\ref{fig:usernum_C}.

% \subsection{Comparison with Doppler-based Method}
% We have discussed the drawbacks of the Doppler-based method in Section~\ref{sec:Related_Work}. In this section, we give a intuitive comparison between the Doppler-based method and FADE on same data sets FALL DS2, FALL DS3 and part of FALL DS1. 

% As shown in Figure~\ref{fig:IMMvsCNN}, the performance of the Doppler-based approach deteriorates when the number of users in the test scene increase, while FADE still maintains a high-level performance. Though the Doppler-based approach and FADE achieve the same recall rate in the single user scenario, the precision the of Doppler-based approach drops due to the incompleteness of the training set (Doppler data varies in different situations).

% \begin{figure}
%   \centering
%   \includegraphics[width=0.95\linewidth]{./figure/IMMvsCNN.pdf}
%   \caption{System performance comparison between FADE and Doppler-based method}
%   \label{fig:IMMvsCNN}
%   \vspace{-3mm}
% \end{figure}

\section{Conclusion}\label{sec:conclusion}

In this study, we use a mmWave radar sensor for fall detection because of its advantages, such as privacy-compliant, device-free, sensitive to motions, and so on. We make an assumption that the motion model of the human torso can be described as a mix of CV and CA models. Then, we introduce the IMM algorithm to estimate the user's acceleration and the probabilities of different motion models as the environment-dependent features. The experimental results show FADE can reach 0.95 F1 score of fall detection in a couple of datasets. Besides, the time cost of FADE is acceptable which offers a chance to future smart home and health monitoring.

\addtolength{\textheight}{0cm}   % This command serves to balance the column lengths
                                  % on the last page of the document manually. It shortens
                                  % the textheight of the last page by a suitable amount.
                                  % This command does not take effect until the next page
                                  % so it should come on the page before the last. Make
                                  % sure that you do not shorten the textheight too much.

%%%%%%%%%%%%%%%%%%%%%%%%%%%%%%%%%%%%%%%%%%%%%%%%%%%%%%%%%%%%%%%%%%%%%%%%%%%%%%%%

%%%%%%%%%%%%%%%%%%%%%%%%%%%%%%%%%%%%%%%%%%%%%%%%%%%%%%%%%%%%%%%%%%%%%%%%%%%%%%%%

% \section*{ACKNOWLEDGMENT}
\clearpage
\bibliographystyle{IEEEtran}
\bibliography{reference}

% Generated by IEEEtran.bst, version: 1.14 (2015/08/26)
\begin{thebibliography}{10}
\providecommand{\url}[1]{#1}
\csname url@samestyle\endcsname
\providecommand{\newblock}{\relax}
\providecommand{\bibinfo}[2]{#2}
\providecommand{\BIBentrySTDinterwordspacing}{\spaceskip=0pt\relax}
\providecommand{\BIBentryALTinterwordstretchfactor}{4}
\providecommand{\BIBentryALTinterwordspacing}{\spaceskip=\fontdimen2\font plus
\BIBentryALTinterwordstretchfactor\fontdimen3\font minus
  \fontdimen4\font\relax}
\providecommand{\BIBforeignlanguage}[2]{{%
\expandafter\ifx\csname l@#1\endcsname\relax
\typeout{** WARNING: IEEEtran.bst: No hyphenation pattern has been}%
\typeout{** loaded for the language `#1'. Using the pattern for}%
\typeout{** the default language instead.}%
\else
\language=\csname l@#1\endcsname
\fi
#2}}
\providecommand{\BIBdecl}{\relax}
\BIBdecl

\bibitem{UnitedNations2019}
\BIBentryALTinterwordspacing
{United Nations}, \emph{{World Population Prospects 2019 Volume I:
  Comprehensive Tables}}.\hskip 1em plus 0.5em minus 0.4em\relax United
  Nations, 2019, vol.~I. [Online]. Available:
  \url{https://population.un.org/wpp/Publications/Files/WPP2019{\_}Volume-I{\_}Comprehensive-Tables.pdf}
\BIBentrySTDinterwordspacing

\bibitem{CAMPBELL1981}
\BIBentryALTinterwordspacing
A.~J. CAMPBELL, J.~REINKEN, B.~C. ALLAN, and G.~S. MARTINEZ, ``{FALLS IN OLD
  AGE: A STUDY OF FREQUENCY AND RELATED CLINICAL FACTORS},'' \emph{Age and
  Ageing}, vol.~10, no.~4, pp. 264--270, 1981. [Online]. Available:
  \url{https://academic.oup.com/ageing/article-lookup/doi/10.1093/ageing/10.4.264}
\BIBentrySTDinterwordspacing

\bibitem{WHO2007}
WHO, ``{WHo Global report on falls Prevention in older Age PAGE},'' World
  Health Organization, Tech. Rep., 2007.

\bibitem{Seketa2021}
\BIBentryALTinterwordspacing
G.~{\v{S}}eketa, L.~Pavlakovi{\'{c}}, D.~D{\v{z}}aja, I.~Lackovi{\'{c}}, and
  R.~Magjarevi{\'{c}}, ``{Event-Centered Data Segmentation in
  Accelerometer-Based Fall Detection Algorithms},'' \emph{Sensors}, vol.~21,
  no.~13, p. 4335, jun 2021. [Online]. Available:
  \url{https://www.mdpi.com/1424-8220/21/13/4335}
\BIBentrySTDinterwordspacing

\bibitem{Lee2015}
\BIBentryALTinterwordspacing
J.~K. Lee, S.~N. Robinovitch, and E.~J. Park, ``{Inertial Sensing-Based
  Pre-Impact Detection of Falls Involving Near-Fall Scenarios},'' \emph{IEEE
  Transactions on Neural Systems and Rehabilitation Engineering}, vol.~23,
  no.~2, pp. 258--266, mar 2015. [Online]. Available:
  \url{https://ieeexplore.ieee.org/document/6905812/}
\BIBentrySTDinterwordspacing

\bibitem{Stone2015}
\BIBentryALTinterwordspacing
E.~E. Stone and M.~Skubic, ``{Fall Detection in Homes of Older Adults Using the
  Microsoft Kinect},'' \emph{IEEE Journal of Biomedical and Health
  Informatics}, vol.~19, no.~1, pp. 290--301, jan 2015. [Online]. Available:
  \url{https://ieeexplore.ieee.org/document/6774430/}
\BIBentrySTDinterwordspacing

\bibitem{Nunez-Marcos2017}
\BIBentryALTinterwordspacing
A.~N{\'{u}}{\~{n}}ez-Marcos, G.~Azkune, and I.~Arganda-Carreras,
  ``{Vision-Based Fall Detection with Convolutional Neural Networks},''
  \emph{Wireless Communications and Mobile Computing}, vol. 2017, p. 9474806,
  2017. [Online]. Available: \url{https://doi.org/10.1155/2017/9474806}
\BIBentrySTDinterwordspacing

\bibitem{Rougier2011}
C.~Rougier, J.~Meunier, A.~St-Arnaud, and J.~Rousseau, ``{Robust Video
  Surveillance for Fall Detection Based on Human Shape Deformation},''
  \emph{IEEE Transactions on Circuits and Systems for Video Technology},
  vol.~21, no.~5, pp. 611--622, 2011.

\bibitem{Auvinet2011}
E.~Auvinet, F.~Multon, A.~Saint-Arnaud, J.~Rousseau, and J.~Meunier, ``{Fall
  Detection With Multiple Cameras: An Occlusion-Resistant Method Based on 3-D
  Silhouette Vertical Distribution},'' \emph{IEEE Transactions on Information
  Technology in Biomedicine}, vol.~15, no.~2, pp. 290--300, 2011.

\bibitem{Tomii2012}
\BIBentryALTinterwordspacing
S.~Tomii and T.~Ohtsuki, ``{Falling detection using multiple doppler
  sensors},'' in \emph{2012 IEEE 14th International Conference on e-Health
  Networking, Applications and Services (Healthcom)}.\hskip 1em plus 0.5em
  minus 0.4em\relax IEEE, oct 2012, pp. 196--201. [Online]. Available:
  \url{http://ieeexplore.ieee.org/document/6379404/}
\BIBentrySTDinterwordspacing

\bibitem{Karsmakers2012}
P.~Karsmakers, T.~Croonenborghs, M.~Mercuri, D.~Schreurs, and P.~Leroux,
  ``{Automatic in-door fall detection based on microwave radar measurements},''
  \emph{European Microwave Week 2012: "Space for Microwaves", EuMW 2012,
  Conference Proceedings - 9th European Radar Conference, EuRAD 2012}, pp.
  202--205, 2012.

\bibitem{Wu2013}
\BIBentryALTinterwordspacing
M.~Wu, X.~Dai, Y.~D. Zhang, B.~Davidson, M.~G. Amin, and J.~Zhang, ``{Fall
  Detection Based on Sequential Modeling of Radar Signal Time-Frequency
  Features},'' in \emph{2013 IEEE International Conference on Healthcare
  Informatics}, no.~1.\hskip 1em plus 0.5em minus 0.4em\relax IEEE, sep 2013,
  pp. 169--174. [Online]. Available:
  \url{http://ieeexplore.ieee.org/document/6680475/}
\BIBentrySTDinterwordspacing

\bibitem{Su2015}
\BIBentryALTinterwordspacing
B.~Y. Su, K.~C. Ho, M.~J. Rantz, and M.~Skubic, ``{Doppler Radar Fall Activity
  Detection Using the Wavelet Transform},'' \emph{IEEE Transactions on
  Biomedical Engineering}, vol.~62, no.~3, pp. 865--875, mar 2015. [Online].
  Available: \url{http://ieeexplore.ieee.org/document/6945894/}
\BIBentrySTDinterwordspacing

\bibitem{Sadreazami2019}
\BIBentryALTinterwordspacing
H.~Sadreazami, M.~Bolic, and S.~Rajan, ``{CapsFall: Fall Detection Using
  Ultra-Wideband Radar and Capsule Network},'' \emph{IEEE Access}, vol.~7, pp.
  55\,336--55\,343, 2019. [Online]. Available:
  \url{https://ieeexplore.ieee.org/document/8703827/}
\BIBentrySTDinterwordspacing

\bibitem{Jin2019}
\BIBentryALTinterwordspacing
F.~Jin, R.~Zhang, A.~Sengupta, S.~Cao, S.~Hariri, N.~K. Agarwal, and S.~K.
  Agarwal, ``{Multiple Patients Behavior Detection in Real-time using mmWave
  Radar and Deep CNNs},'' in \emph{2019 IEEE Radar Conference
  (RadarConf)}.\hskip 1em plus 0.5em minus 0.4em\relax IEEE, apr 2019, pp.
  1--6. [Online]. Available:
  \url{https://ieeexplore.ieee.org/document/8835656/}
\BIBentrySTDinterwordspacing

\bibitem{Takabatake2019}
\BIBentryALTinterwordspacing
W.~Takabatake, K.~Yamamoto, K.~Toyoda, T.~Ohtsuki, Y.~Shibata, and A.~Nagate,
  ``{FMCW Radar-Based Anomaly Detection in Toilet by Supervised Machine
  Learning Classifier},'' in \emph{2019 IEEE Global Communications Conference
  (GLOBECOM)}, no.~i.\hskip 1em plus 0.5em minus 0.4em\relax IEEE, dec 2019,
  pp. 1--6. [Online]. Available:
  \url{https://ieeexplore.ieee.org/document/9014123/}
\BIBentrySTDinterwordspacing

\bibitem{Wang2020}
\BIBentryALTinterwordspacing
B.~Wang, L.~Guo, H.~Zhang, and Y.-X. Guo, ``{A Millimetre-Wave Radar-Based Fall
  Detection Method Using Line Kernel Convolutional Neural Network},''
  \emph{IEEE Sensors Journal}, vol.~20, no.~22, pp. 13\,364--13\,370, nov 2020.
  [Online]. Available: \url{https://ieeexplore.ieee.org/document/9133594/}
\BIBentrySTDinterwordspacing

\bibitem{Ma2020}
\BIBentryALTinterwordspacing
L.~Ma, M.~Liu, N.~Wang, L.~Wang, Y.~Yang, and H.~Wang, ``{Room-Level Fall
  Detection Based on Ultra-Wideband (UWB) Monostatic Radar and Convolutional
  Long Short-Term Memory (LSTM)},'' \emph{Sensors}, vol.~20, no.~4, p. 1105,
  feb 2020. [Online]. Available: \url{https://www.mdpi.com/1424-8220/20/4/1105}
\BIBentrySTDinterwordspacing

\bibitem{Hanifi2021}
\BIBentryALTinterwordspacing
K.~Hanifi and M.~{Elif Karsligil}, ``{Elderly Fall Detection with Vital Signs
  Monitoring Using CW Doppler Radar},'' \emph{IEEE Sensors Journal}, vol.~XX,
  no.~XX, pp. 1--1, 2021. [Online]. Available:
  \url{https://ieeexplore.ieee.org/document/9429253/}
\BIBentrySTDinterwordspacing

\bibitem{Maitre2021}
\BIBentryALTinterwordspacing
J.~Maitre, K.~Bouchard, and S.~Gaboury, ``{Fall Detection With UWB Radars and
  CNN-LSTM Architecture},'' \emph{IEEE Journal of Biomedical and Health
  Informatics}, vol.~25, no.~4, pp. 1273--1283, apr 2021. [Online]. Available:
  \url{https://ieeexplore.ieee.org/document/9212552/}
\BIBentrySTDinterwordspacing

\bibitem{Jin2020}
\BIBentryALTinterwordspacing
F.~Jin, A.~Sengupta, and S.~Cao, ``{mmFall: Fall Detection Using 4-D mmWave
  Radar and a Hybrid Variational RNN AutoEncoder},'' \emph{IEEE Transactions on
  Automation Science and Engineering}, pp. 1--13, 2020. [Online]. Available:
  \url{https://ieeexplore.ieee.org/document/9305931/}
\BIBentrySTDinterwordspacing

\bibitem{Wang2020a}
\BIBentryALTinterwordspacing
X.~Wang, J.~Ellul, and G.~Azzopardi, ``{Elderly Fall Detection Systems: A
  Literature Survey},'' \emph{Frontiers in Robotics and AI}, vol.~7, jun 2020.
  [Online]. Available:
  \url{https://www.frontiersin.org/article/10.3389/frobt.2020.00071/full}
\BIBentrySTDinterwordspacing

\bibitem{Blom1988}
\BIBentryALTinterwordspacing
H.~Blom and Y.~Bar-Shalom, ``{The interacting multiple model algorithm for
  systems with Markovian switching coefficients},'' \emph{IEEE Transactions on
  Automatic Control}, vol.~33, no.~8, pp. 780--783, 1988. [Online]. Available:
  \url{http://ieeexplore.ieee.org/document/1299/}
\BIBentrySTDinterwordspacing

\bibitem{Amin2016}
\BIBentryALTinterwordspacing
M.~G. Amin, Y.~D. Zhang, F.~Ahmad, and K.~D. Ho, ``{Radar Signal Processing for
  Elderly Fall Detection: The future for in-home monitoring},'' \emph{IEEE
  Signal Processing Magazine}, vol.~33, no.~2, pp. 71--80, mar 2016. [Online].
  Available: \url{http://ieeexplore.ieee.org/document/7426551/}
\BIBentrySTDinterwordspacing

\bibitem{Tian2018}
Y.~Tian, G.-H. Lee, H.~He, C.-Y. Hsu, and D.~Katabi, ``{RF-Based Fall
  Monitoring Using Convolutional Neural Networks},'' \emph{Proceedings of the
  ACM on Interactive, Mobile, Wearable and Ubiquitous Technologies}, vol.~2,
  no.~3, pp. 1--24, 2018.

\bibitem{Noury2016}
\BIBentryALTinterwordspacing
N.~Noury, J.~Poujaud, P.~Cousin, and N.~Poujaud, ``{Biomechanical analysis of a
  fall: Velocities at impact},'' in \emph{2016 38th Annual International
  Conference of the IEEE Engineering in Medicine and Biology Society (EMBC)},
  vol. 2016-Octob.\hskip 1em plus 0.5em minus 0.4em\relax IEEE, aug 2016, pp.
  561--565. [Online]. Available:
  \url{http://ieeexplore.ieee.org/document/7590764/}
\BIBentrySTDinterwordspacing

\bibitem{Martinez-Villasenor2019}
\BIBentryALTinterwordspacing
L.~Mart{\'{i}}nez-Villase{\~{n}}or, H.~Ponce, J.~Brieva, E.~Moya-Albor,
  J.~N{\'{u}}{\~{n}}ez-Mart{\'{i}}nez, and C.~Pe{\~{n}}afort-Asturiano,
  ``{UP-Fall Detection Dataset: A Multimodal Approach},'' \emph{Sensors},
  vol.~19, no.~9, p. 1988, apr 2019. [Online]. Available:
  \url{https://www.mdpi.com/1424-8220/19/9/1988}
\BIBentrySTDinterwordspacing

\bibitem{BarShalom1989}
Y.~{Bar Shalom}, K.~C. Chang, and H.~A. Blom, ``{Tracking a Maneuvering Target
  Using Input Estimation Versus the Interacting Multiple Model Algorithm},''
  \emph{IEEE Transactions on Aerospace and Electronic Systems}, vol.~25, no.~2,
  pp. 296--300, 1989.

\bibitem{Challa2011}
\BIBentryALTinterwordspacing
S.~Challa, M.~R. Morelande, D.~Musicki, and R.~J. Evans, \emph{{Fundamentals of
  Object Tracking}}.\hskip 1em plus 0.5em minus 0.4em\relax Cambridge:
  Cambridge University Press, 2011. [Online]. Available:
  \url{http://ebooks.cambridge.org/ref/id/CBO9780511975837}
\BIBentrySTDinterwordspacing

\end{thebibliography}
\newpage

% \begin{thebibliography}{99}

% \end{thebibliography}

\end{document}